\documentclass[11pt, oval, oneside]{article} 

\usepackage[utf8]{inputenc}
\usepackage[T1]{fontenc}
\usepackage{amsmath}
\usepackage{booktabs}
\usepackage{amssymb}
\usepackage{commath}
\usepackage{graphicx}
\usepackage{subcaption}
\usepackage{lineno}
\usepackage{xcolor}
\usepackage{hyperref}
\usepackage[margin=1in]{geometry}    
\usepackage{authblk}
\usepackage{float}
\usepackage{setspace}

\begingroup
\catcode`&=12
\gdef\aap{A&A}
\endgroup

\begingroup
\catcode`&=12
\gdef\araa{ARA&A}
\endgroup

\newcommand{\pasa}{PASA}
\newcommand{\prl}{Phys.~Rev.~Lett.}
\newcommand{\mnras}{MNRAS}
\newcommand{\nat}{Nature}
\newcommand{\apj}{ApJ}
\newcommand{\apjl}{ApJL}
\newcommand{\apjs}{ApJS}

\newcommand{\prd}{Phys. Rev. D}
\newcommand{\pasp}{PASP}

\newcommand{\JournalTitle}[1]{\textit{#1}}

\newcommand{\gendirect}{\textsc{GenDIReCT}}
\newcommand{\bilby}{\textsc{bilby}}

\newcommand{\ehtim}{\textsc{eht-imaging}}
\newcommand{\rex}{\textsc{REx}}
\newcommand{\difmap}{\textsc{DIFMAP}}
\newcommand{\themis}{\textsc{THEMIS}}
\newcommand{\comrade}{\textsc{Comrade}}

\newcommand{\clean}{\textsc{CLEAN}}
\newcommand{\doghit}{\textsc{DoG-HiT}}
\newcommand{\smili}{\textsc{SMILI}}
\newcommand{\hops}{\textsc{EHT-HOPS}}
\newcommand{\dynesty}{\textsc{dynesty}}
\newcommand{\ipole}{\textsc{ipole}}

\newcommand{\mes}{M87$^{*}$}

\newcommand{\xsrg}{\textit{xsringauss}}
\newcommand{\cagn}{3C~279}
\newcommand{\cena}{Centaurus~A}

\title{Independent interferometric observable confirms black hole shadow in \mes}
\date{}

\author[1,*]{Samuel Lai}
\author[1]{Nithyanandan Thyagarajan}
\author[1,2]{O. Ivy Wong}
\author[3]{Foivos Diakogiannis}

\affil[1]{Space \& Astronomy, Commonwealth Scientific and Industrial Research Organisation (CSIRO), P. O. Box 1130, Bentley, WA 6102, Australia}
\affil[2]{International Centre for Radio Astronomy Research, The University of Western Australia, Crawley, WA 6009, Australia}
\affil[3]{Technology, Commonwealth Scientific and Industrial Research Organisation (CSIRO), Kensington, WA 6151, Australia}
\affil[*]{Corresponding author: samuel.lai@csiro.au}

\begin{document}
\maketitle

{\bf 
A new epoch for direct visualisation of accretion dynamics near a black hole's event horizon began when the Event Horizon Telescope Collaboration (EHTC) synthesised an Earth-sized aperture with millimeter radio interferometry and published the first reconstructed image\cite{EHT_2019_Array, EHT_2019_Data, EHT_2019_Imaging, EHT_2019_ShadowMass} of a black hole shadow at the core of radio galaxy Messier 87 (M87). The image appeared as a ring-like structure, representing the gravitationally lensed and redshifted emission around the central supermassive black hole, \mes, located at the galactic nucleus. However, application of various data analysis and imaging heuristics resulted in divergent representations of \mes\ fine-scale features\cite{EHT_2019_Imaging, Arras_2022, Carilli_2022, Lockhart_2022}, with some external analyses even challenging the fundamental ring-like morphology of the source\cite{Miyoshi_2022_M87, Miyoshi_2024}. Such difficulties arise as a consequence of the inherent sparsity of the very-long baseline interferometry (VLBI) measurements, rendering interferometric image reconstruction an inadequately constrained problem. Moreover, the solution is sensitive to instrument calibration, post-processing algorithms, and imaging hyperparameters. 
Here, we apply a newly-developed independent imaging algorithm to the \mes\ dataset, which is free from prescriptive morphological priors and founded on observables immune to instrument calibration errors.
We reconstruct the black hole shadow image, and unequivocally reproduce a ring-like morphology, with angular diameter $42.2\pm 1.5$ microarcseconds corresponding to a black hole mass of $6.5\pm 0.5$ billion solar masses assuming nominal scaling parameters, observed at an inclination angle of $\sim20^{\circ}$ from the approaching jet.  
}

\smallskip

The \mes\ supermassive black hole, with a mass of $M_{\rm BH} = (3-7)\times 10^9\,M_{\odot}$\cite{Gebhardt_2009, Gebhardt_2011, Walsh_2013} located at a distance of $D=16.8\,{\rm Mpc}$\cite{Blakeslee_2009, Bird_2010}, offers a unique window into horizon-scale general relativistic physics. Its appearance on event horizon scales is expected, based on general relativistic predictions, to be dominated by a central shadow, which for a non-rotating Schwarzschild black hole, is characterized by a critical impact parameter that is proportional to the black hole mass\cite{Ames_1968_impactparam}
as $3\sqrt{3}GM_{\rm{BH}}/(c^2 D)$, where $G$ is the gravitational constant and $c$ is the speed of light. To observe with a spatial resolution capable of detailing the \mes\ black hole shadow requires unprecedented microarcsecond-scale angular resolution, roughly equivalent to the angle subtended by the width of a strand of human hair over 1000~km away. Achieving this feat requires a combination of millimeter wavelengths ($\lambda \sim 1 \,{\rm mm}$) and an effective aperture size comparable to the Earth's diameter.

Such a synthesis of an Earth-sized aperture was realised through VLBI techniques, by correlating the signals from a global network of independent radio observatories, collectively referred to as the Event Horizon Telescope (EHT)\cite{Doeleman_2009, EHT_2019_Array}. As the Earth rotates, the separation vectors between pairs of geographical stations sweep out elliptical arcs that incrementally fill the Earth-sized synthetic aperture. While this technique yields the requisite angular resolution to image the \mes\ shadow, the resulting virtual aperture remains significantly unfilled and lacks the sensitivity and reconstruction fidelity of a filled aperture. The data sparsity problem is further compounded by the requirement of calibrating rapid, station-based atmospheric phase and amplitude fluctuations on continental-scale separations, which often occur on timescales shorter than the coherence time of the observations.

These inherent challenges render image reconstruction from VLBI and EHT datasets an ill-posed problem, highly sensitive to instrument calibration and post-processing heuristics. Current algorithms regularly employ manually-tuned and independently weighted data terms and regularisations to bias the solution toward plausible morphologies, which can inadvertently steer the reconstruction toward preferred solutions. The consequences are divergent representations of \mes's fine-scale features, such as the surface brightness distribution and hotspot localization\cite{EHT_2019_Imaging, Arras_2022, Carilli_2022, Broderick_2022, Lockhart_2022, Muller_2024_closureTraces}. Using traditional approaches, including \clean-based algorithms or hybrid mapping with self-calibration\cite{Readhead_1978, Readhead_1980, Cornwell_1981, Pearson_1984_SelfCal, Cornwell_1999}, independent analyses showed that the resulting reconstruction is sensitive to methodological and hyperparameter selections, such as image initialisation\cite{Carilli_2022} or adopted field of view\cite{Miyoshi_2022_M87}, with some independent analyses demonstrating reconstructions without a clear ring-like structure\cite{Miyoshi_2022_M87, Miyoshi_2024}, suggesting that if the field of view extends to circular diameters beyond $120\,\mu{\rm as}$, a single elongated bright spot solution is favoured over a ring\cite{Miyoshi_2022_M87}. 

To resolve the tension around the \mes\ results and advance beyond methodological uncertainties, we transition from traditional approaches toward a more autonomous and objective framework. The independence of our method is primarily grounded on our use of robust interferometric observables, called closure invariants\cite{Jennison_1958, Twiss_1960, Thyagarajan_2022_CI,Samuel_2022}, which capture information on source morphology that is unaffected by errors in wavefront and instrument calibration. Therefore, closure invariants constitute robust information-bearing observables\footnote{Closure invariants are also insensitive to absolute position and absolute flux scale, but these are not inherent morphological parameters.} in radio interferometry, underpinning high-fidelity morphology reconstructions\cite{Chael_2018_ehtim, Thyagarajan_2024_Lucas, Lai_2025_DIReCT, Lai_2025_GenDIReCT}. In contrast to the disparate closure phases\cite{Jennison_1958} and closure amplitudes\cite{Twiss_1960} that have traditionally served as such robust observables, the novel closure invariants formalism\cite{Thyagarajan_2022_CI, Samuel_2022} that we employ provides a unified framework that carries equivalent information while eliminating the requirement for heterogeneous treatment typically observed in the literature. By leveraging the rich calibration-independent, pure morphological information carried by closure invariants, we can bypass conventional techniques which require sensitive calibration tuning and observer-driven priors.

\subsection*{Novel interferometric imaging}
To take full advantage of the novel closure formalism, we develop and apply the Generative Deep learning Image Reconstruction with Closure Terms algorithm (\gendirect)\cite{Lai_2025_GenDIReCT}, trained on images unrelated to astronomy and designed to generatively produce data-driven image reconstructions using only closure invariants. Unlike other machine-learning based radio interferometric imaging algorithms, our algorithm is unique in employing closure invariants to condition a latent diffusion model. \gendirect\ has undergone extensive testing on both synthetic\cite{Lai_2025_GenDIReCT} and real data\cite{Lai_2026_3C279CenA}, and has been demonstrated to be capable of consistently recovering diverse morphological structures, without manual intervention, prescriptive priors, or parameter tuning (additional detail in \hyperref[methods:gendirect]{Methods}).

For the first time, closure invariants from the new formalism are used to image \mes\ through our \gendirect\ algorithm, using the publicly available $\sim1.3\,{\rm mm}$ EHT measurements from four observation epochs (April 5, 6, 10, 11) from 2017\cite{EHT_2019_Array}, two epochs (April 21, 25) from 2018\cite{EHT_2024_M87_PerShad}, and two epochs (April 13, 18) from 2021\cite{EHT_2025_obs2021}. We set multiple fields of view for the reconstruction ($128\times128$, $225\times225$, and $450\times450\,\mu\rm{as}^2$), where each combination of dataset and field of view is an independent model and acts as an indirect prior on the maximum source size. For each of these models, we aggregate a large population of generative reconstructions into a single representative median image, comparing them to the EHTC fiducial reconstruction which is reproduced using the public \ehtim\ pipeline\footnote{\href{https://github.com/eventhorizontelescope/2019-D01-02}{https://github.com/eventhorizontelescope/2019-D01-02}}. We also derive a composite image from the pixel-wise median of all reconstructions of each epoch across all fields of view aligned to the single reconstruction with the best data fidelity score. 

\begin{figure}[h]
\begin{center}
\resizebox{15cm}{!}{\includegraphics[trim={0 0 0 0},clip]{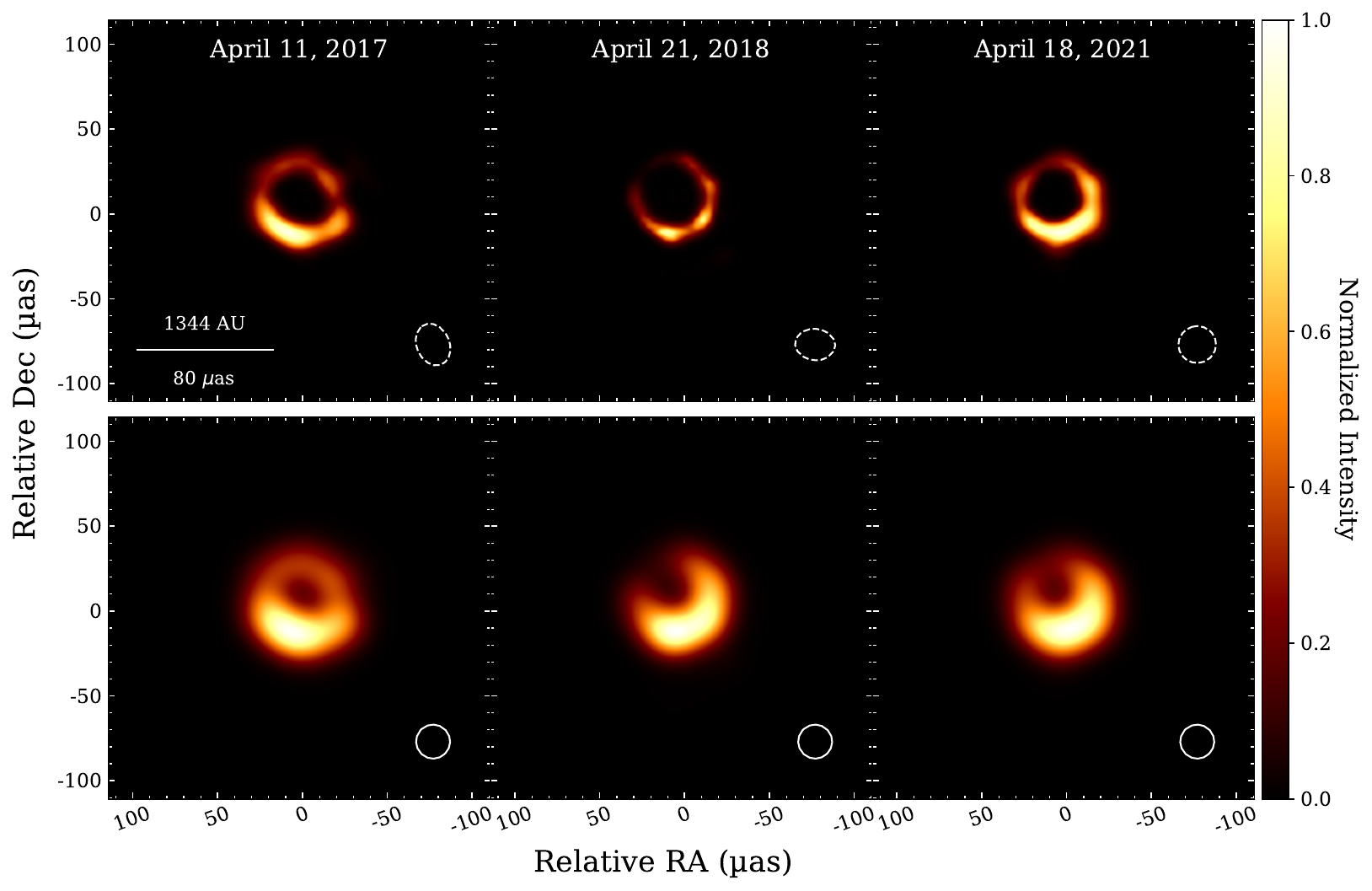}}
\end{center}
\caption{\textbf{Composite image reconstructions} produced with the April 11, 2017 (left), April 21, 2018 (middle), and April 18, 2021 (right) datasets. The dashed unfilled ellipses represent the full-width half-maxima of the synthesised point-spread-functions. The bottom row illustrates the composite reconstructions convolved with a 20 $\mu$as circular beam, comparable with fiducial EHTC image reconstructions of \mes. Image reconstructions in the top row are unconvolved. Here, intensity is illustrated in a normalised linear scale, where the compact flux density of \mes\ is constrained by other means to be between $0.4-1.0$ Jy\cite{EHT_2019_Imaging, EHT_2024_M87_PerShad, EHT_2025_obs2021}.}
\label{fig:final-composites}
\end{figure}

All datasets from independent observation epochs convincingly produce compact, ring-like structure on $\sim40\,\mu{\rm as}$-scale. On the highest quality dataset of each year (April 11, 2017\footnote{April 11 has the highest signal-to-noise ratio in closure invariants of all datasets in 2017. The EHTC identified the April 11 dataset as exhibiting the most stable a priori amplitude calibration, and the most shared coverage with the \cagn\ calibrator source, using it as the fiducial dataset\cite{EHT_2019_Imaging}.}, April 21, 2018, and April 18, 2021), our median reconstruction is a compact circularised ring of consistent orientation and diameter, independent of the reconstruction field-of-view window and brightened in the south or south-west direction. The appearance of the ring is most consistent with an observed inclination angle of $\sim20^{\circ}$ from the approaching jet. We present the composites for our best reconstructions from these three datasets in Figure \ref{fig:final-composites}.

The consistency of the source size ($\sim40\,\mu{\rm as}$) across all fields of view is strong confirmation of a compact source. Under the most-constrained $128\times128\,\mu{\rm as}^2$ field of view, the median reconstructed image exhibits the highest image-domain similarity score with fiducial EHTC reconstructions and the performance on data adherence metrics is competitive with those of larger fields of view, supporting the prior expectation that extended emission of the \mes\ jet is unconstrained by the EHT datasets\cite{EHT_2019_Imaging, EHT_2024_M87_PerShad}. On some infrequent reconstructions of epochs with lower-quality datasets suffering from even sparser aperture coverage\cite{EHT_2019_Imaging}, lower signal-to-noise ratio, or various observation deficiencies\cite{EHT_2024_M87_PerShad}, the ring-like morphology can become less clear, as noted by independent studies\cite{Miyoshi_2022_M87}. Consequently, the suggested sensitivity of image morphology to the assumed field of view reported in independent re-analyses and demonstrated for the April 5, 2017 dataset\cite{Miyoshi_2022_M87} is possibly an artifact of insufficient aperture coverage or low quality data. In our work, when higher quality datasets are considered, the sensitivity to field of view disappears, contrary to predictions based on the traditional hybrid mapping approach\cite{Miyoshi_2022_M87}. Our work thus yields a robust, independent verification of the ring-like geometry characteristic of the \mes\ black hole shadow image.

In contrast to regularised maximum likelihood methods in the publicly available imaging pipelines used by the EHTC on \mes, our method does not require manually prescribing an image initialisation. For instance, the \ehtim\ pipeline applied on \mes\ is initialised with a $40\,\mu{\rm as}$ full-width at half-maximum Gaussian, which serves as a powerful prior on the characteristic source size. These methods typically exhibit significant sensitivity to the starting model\cite{Carilli_2022}. When the \ehtim\ pipeline is initialised with Gaussian priors larger than $40\,\mu{\rm as}$ (i.e. $80-120\,\mu{\rm as}$), the resulting image reconstructions exhibit phantom repetition or aliasing artifacts, which is a common issue for VLBI imaging algorithms\cite{Chael_2018_ehtim, Muller_2023_MOEAD}.

It is significant, therefore, that we consistently recover a single, compact source morphology in the absence of manual prescriptions. When our model is trained on different fields of view, the model learns an implicit statistical prior from its diverse training library conditioned on the input interferometric observables, namely, closure invariants. The convergence toward a stable, singular ring structure across varying fields of view indicates that the our model is successfully inferring the intrinsic source geometry from the data itself, rather than operating within the constraints of a manually-selected initialisation. \gendirect's autonomy and capacity for imaging directly from calibration-independent observables represents a major step toward objective blind imaging in the challenging data-sparse VLBI regime. By unequivocally recovering the ring-like structure of \mes\ across multiple datasets through this approach, we provide a rigorous cross-validation of published EHTC results. 

\subsection*{Morphological analysis}

We quantitatively demonstrate consistency with the EHTC image reconstructions by measuring ring characteristics from our reconstructed images and comparing them to published parameters. We apply an independent implementation of the image-domain feature extraction method utilised by the EHTC\cite{EHT_2019_Imaging} to measure ring properties from reconstructed event-horizon-scale images, including ring diameter, width, orientation (measured counter-clockwise from North), azimuthal asymmetry and central brightness fraction. The measured properties from our median reconstructions of all datasets are presented in Table \ref{tab:morph-features}. Each cell represents the mean parameter measured from independent reconstructions on all three fields of view.

\begin{figure}[h]
\begin{center}
\resizebox{13cm}{!}{\includegraphics[trim={0 0 0 0},clip]{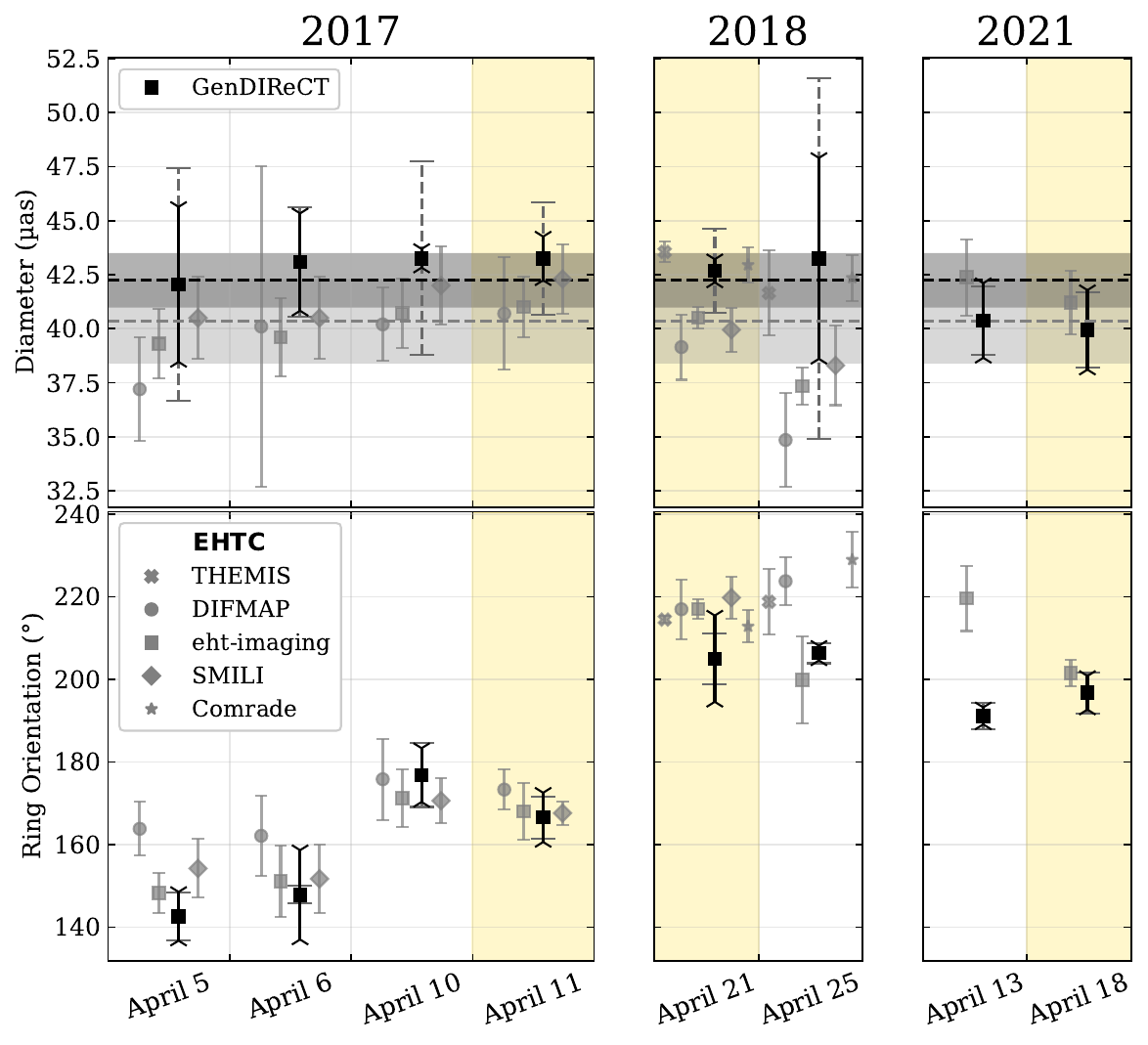}}
\end{center}
\caption{\textbf{Measured ring diameter and orientation.} Comparison of ring diameter and orientation measured from image-domain feature extraction applied on our reconstructions with values from the EHTC, separated by observation epoch. EHTC measurements on each epoch correspond to different imaging algorithms (from left-to-right: \themis, \difmap, \ehtim, \smili, \comrade). Our measurements are presented with two estimates of uncertainty: the dashed line delineates the dispersion 
(along azimuthal axis for diameter and along radial axis for orientation)
in the image-domain feature extraction method and the solid line with triad cap indicates the standard deviation across fields of view, which is unique to our method. For the diameter, we illustrate the mean and 1$\sigma$ standard deviation for EHTC methods and our method with the dashed line and sheath. The highest quality dataset of each year is highlighted in gold.
}
\label{fig:ringdiameter}
\end{figure}

The measured ring orientation and diameter remain relatively stable between \gendirect\ and EHTC imaging methods. In Figure \ref{fig:ringdiameter}, we compare our image-domain measurements derived from our reconstructions to the EHTC analogue.  There is a clear counter-clockwise trend in the ring orientation angle as measured by all methods between the April 5 and 6 reconstructions with April 10 and 11 in 2017, as well as between the 2017 and 2018/2021 observations. However, the evolution within 2017 may be caused by uncertainty in the azimuthal structure brought about by the sparse aperture coverage of observations taken on April 5 and 6. Nevertheless, the evolution in ring orientation between the 2017 and 2018 observations appears to be significant, attributable to the turbulent nature of black hole accretion\cite{EHT_2025_PerShadowII}. In particular, the observed ring orientation in 2018 and 2021 are closer in alignment with the large-scale forward jet of \mes, compared to that in 2017.

Conversely, the measured ring width and central brightness fraction are heavily influenced by methodological differences as seen in the variation between imaging algorithms. For instance, reconstructions from the \difmap\ algorithm\cite{Shepherd_2011_DIFMAP} are convolved with a $20\,\mu{\rm as}$ circular beam, which inflates the ring width compared to regularised maximum likelihood methods and our algorithm. Likewise, the azimuthal asymmetry parameter is unstable between different days and different methods, which is indicative of sensitivity to variation in the aperture coverage. Therefore, the ring width and central brightness fraction likely do not reflect intrinsic source properties.

\begin{figure}[h]
\begin{center}
\resizebox{18cm}{!}{\includegraphics[trim={0 0 0 0},clip]{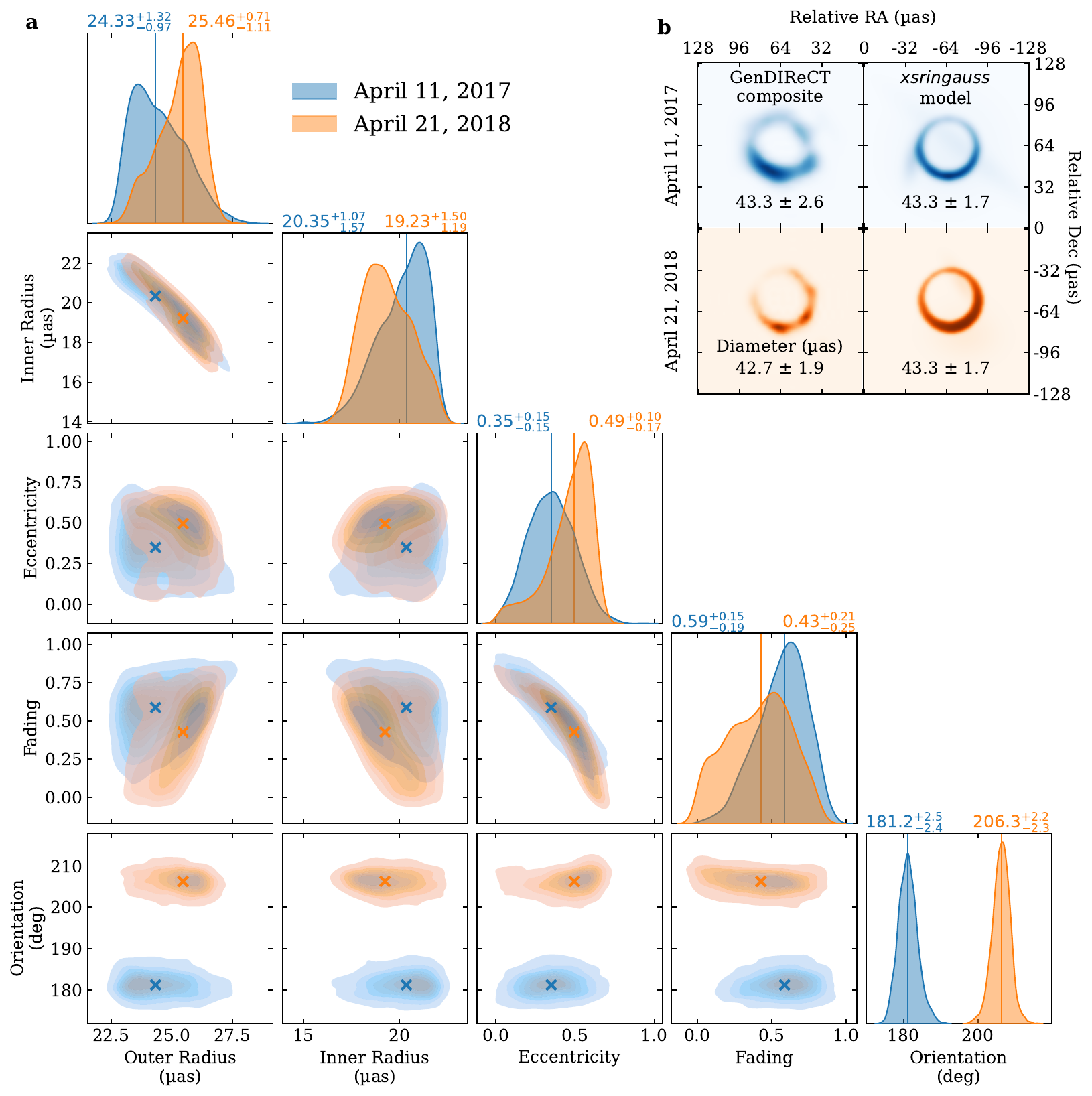}}
\end{center}
\caption{\textbf{Posterior covariance from \xsrg\ geometric modelling} on the April 11, 2017 and April 21, 2018 datasets. Panels show the one-dimensional and two-dimensional marginalised posterior distributions and median values (\textbf{a}) for the outer radius, inner radius, eccentricity, fading parameter, and orientation. The composite images created from all \gendirect\ reconstructions across all fields of view are presented in comparison to the fitted median \xsrg\ sky models (\textbf{b}). The diameter is measured with image-domain feature extraction and the corresponding uncertainty is from azimuthal dispersion rather than the sample variance across multiple fields of view. 
}
\label{fig:corner-main}
\end{figure}

To further cross-validate the measured ring parameters, we employ a complementary Bayesian forward-modelling framework to fit the observed closure invariants using a superposition of primitive geometric shapes with analytic visibilities. With the ring-like structure firmly established by our independent, calibration-insensitive, and non-parametric imaging algorithm, we proceed to construct our geometric composite using the \xsrg\cite{EHT_2019_ShadowMass} model as a foundation. We include one centralised large Gaussian and two elliptical nuisance Gaussians with independent spatial offsets to help capture non-ring short-spacing flux and extraneous emission. Using the \dynesty\ nested sampling algorithm\cite{dynesty}, we perform parameter estimation for the \xsrg\ model for each epoch. Figure \ref{fig:corner-main} shows the posterior covariance from geometric modelling of the April 11, 2017 and April 21, 2018 datasets, where the one-dimensional and two-dimensional marginalised posterior distributions of the outer radius, inner radius, eccentricity, fading parameter, and orientation are illustrated. We apply the image-domain feature extraction method on the geometric image model derived from the median value of each one-dimensional posterior distribution to measure all characteristic ring parameters. Posterior covariances for all other datasets are provided in \hyperref[sec:methods]{Methods}. In all epochs, we measure a mean diameter and standard deviation of $d = 42.3 \pm 1.1\,\mu{\rm as}$ from forward geometric modelling of closure invariants, which is compatible with the mean $d = 42.2 \pm 1.5\,\mu{\rm as}$ from our non-parametric image reconstructions. 

Inferring the black hole's gravitational radius, and hence its mass, from the measured ring diameter requires independently calibrating the imaging method with synthetic data from simulated models of lensed black hole accretion emission\cite{EHT_2019_ShadowMass}. The prior expectation for the mass of the central black hole in M87 is primarily sourced from high-quality gas\cite{Walsh_2013} and stellar\cite{Gebhardt_2011} dynamical estimates, which measured $M_{\rm gas} = 3.5^{+0.9}_{-0.7} \times10^9\, M_{\odot}$ and $M_{\rm star} = (6.6 \pm0.4)\times10^9\,M_{\odot}$, respectively. Using our measured ring diameter, distance measure of  $16.8^{+0.8}_{-0.7}\,{\rm Mpc}$, and nominal scaling parameters for the gravitational radius, we estimate a black hole mass of $M_{\rm BH} = (6.5 \pm 0.5) \times 10^9\, M_{\odot}$ for \mes.

Our independently measured ring diameters and inferred black hole mass from \gendirect\ imaging and geometric modelling are therefore in excellent agreement with prior stellar dynamical mass measurements\cite{Gebhardt_2011} and with image-domain feature extraction on the imaging methods applied by the EHTC. In contrast, the smaller ring diameter predicted by the black hole mass estimate from gas dynamics would have been challenging to resolve given the EHT beam dimensions of $\sim20\,\mu{\rm as}$. 
Upcoming data from the James Webb Space Telescope Near Infrared Spectrograph integral field unit will provide additional detail on the kinematics within the central arcsecond of the M87 core\cite{AlAmri_2025}, producing finer constraints on the \mes\ black hole mass. 

\begin{figure}[h]
\begin{center}
\resizebox{18cm}{!}{\includegraphics[trim={0 0 0 0},clip]{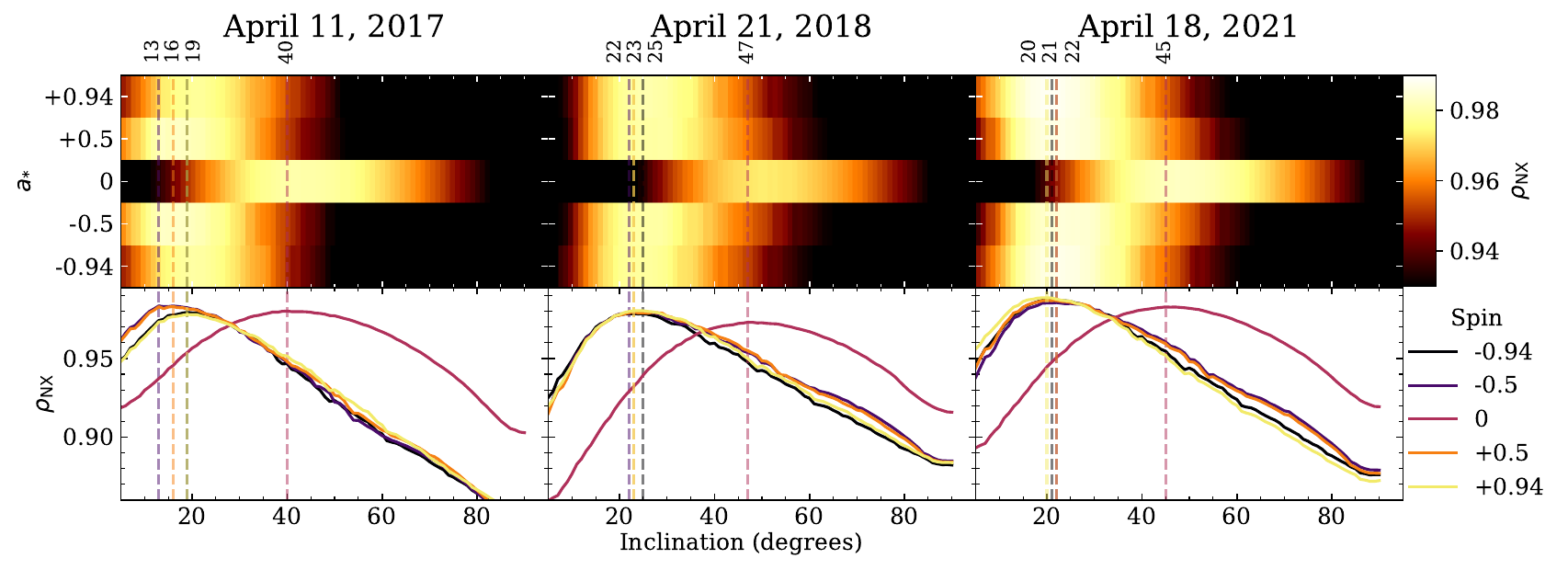}}
\end{center}
\caption{\textbf{Image-domain similarity score across range of black hole spin and observed inclinations} for simulated \mes\ images compared to composite reconstructions for the highest-quality datasets of each year. (Top) Illustration of the covariance in image-domain similarity with black hole spin and inclination. (Bottom) Relationship between inclination and image-domain similarity, separated by black hole spin.
}
\label{fig:spin-inc}
\end{figure}

The appearance of the black hole shadow is also sensitive to the black hole spin, $a_{*}$, and observed inclination angle. We compare our composite \mes\ reconstructions to synthetic 230 GHz images of \mes\ produced by the \ipole\ algorithm\cite{Moscibrodzka_2018_ipole}, which integrates the covariant radiative transfer equation over relativistic photon trajectories across fluid data from general-relativistic magnetohydrodynamic simulations\cite{Dhruv_2025_illinoisGRMHD}. Quantitative analysis using the normalized cross-correlation similarity score, $\rho_{\rm NX}$, comparing beam-convolved synthetic images to image reconstructions, identifies a clear sensitivity to inclination. We present the results based on composites from highest-quality datasets for all years in Figure \ref{fig:spin-inc}. Note that we do not explore inclination angles $>90^{\circ}$ as the resulting synthetic images are comparable to an inversion of the spin direction. While the highest similarity scores favour a non-zero black hole spin, our analysis does not constrain the spin magnitude, consistent with previous analyses with EHT\cite{EHT_2025_PerShadowII} and non-EHT data\cite{Cui_2023}. Altogether, the best-scoring inclination is $\sim20^{\circ}$, and although the best-scoring inclination in the $a_{*}=0$ case is significantly higher, such models are disfavoured a priori by independent considerations of jet power\cite{Reynolds_1996_M87jet, Broderick_2015_M87, Prieto_2016} and precession\cite{Cui_2023}. Consequently, the results from our image-domain comparative analysis are generally consistent with the fiducial $17.2^{\circ}\pm3.3^{\circ}$ jet viewing angle inferred from 43 GHz VLBI observations\cite{Mertens_2016, Walker_2018_17deg}. Future horizon-scale imaging with space-based VLBI offers the potential to resolve the \mes\ photon ring\cite{Johnson_2024_BHEX}, which would enable a direct and precise determination of the black hole spin, currently unconstrained by an image-domain analysis.

\newpage

\begin{table}[h]
\begin{center}
\caption {\label{tab:morph-features} \textbf{Tabulated ring properties} measured from image-domain feature extraction, where each cell represents the mean and standard deviation estimate from our unconvolved reconstructions on all three fields of view. The mean and standard deviation across all epochs for the ring diameter is listed as a summary statistic.
}
\resizebox{\textwidth}{!}{
\begin{tabular}{ll|ccccc}
\hline \hline
Date &  & Ring diameter & Ring width & Orientation & Azimuthal asymmetry & Central brightness fraction \\
& & $2\bar r_{\rm pk}$ [$\mu{\rm{as}}$] & $w_r$ [$\mu${\rm{as}}] & $\eta$ [$^{\circ}$] & $A$ & $f_c$ \\ \hline
2017 & April 5 & $42.0\pm3.6$ & $17\pm2$ & $143\pm6$ & $0.29\pm0.04$ & $0.115\pm0.062$
\\
& April 6 & $43.1\pm2.3$ & $14\pm2$ & $148\pm11$ & $0.33\pm0.04$ & $0.032\pm0.026$
\\
& April 10 & $43.3\pm0.4$ & $22\pm7$ & $177\pm7$ & $0.36\pm0.07$ & $0.135\pm0.066$
\\ 
& April 11 & $43.3\pm1.0$ & $16\pm3$ & $167\pm6$ & $0.20\pm0.01$ & $0.037\pm0.028$
\\ \hline
2018 & April 21 & $42.7\pm0.5$ & $11\pm1$ & $205\pm10$ & $0.31\pm0.01$ & $0.004\pm0.004$
\\
& April 25 & $43.3\pm4.6$ & $36\pm13$ & $206\pm2$ & $0.41\pm0.18$ & $0.270\pm0.135$
\\ \hline
2021 & April 13 & $40.4\pm1.7$ & $17\pm9$ & $191\pm2$ & $0.43\pm0.03$ & $0.002\pm0.001$ \\ 
& April 18 & $39.9\pm1.8$ & $12\pm1$ & $197\pm4$ & $0.26\pm0.01$ & $0.004\pm0.002$
\\ \hline 
Mean & & $42.2\pm 1.5$ & -- & -- & -- & --
\\
\hline \hline
\end{tabular}
}
\end{center}
\end{table}

\newpage

\clearpage

\renewcommand{\figurename}{\textbf{Extended Data Figure}}
\renewcommand{\tablename}{\textbf{Extended Data Table}}
\setcounter{figure}{0}
\setcounter{table}{0}

\section*{Methods} \label{sec:methods}

\subsection*{EHT datasets on \mes}
This work utilises the publicly available network-calibrated datasets of \mes\footnote{\href{https://eventhorizontelescope.org/for-astronomers/data}{https://eventhorizontelescope.org/for-astronomers/data}}, reduced via the \hops\ pipeline, based on observations obtained in 2017\cite{EHT_2019_Data}, 2018\cite{EHT_2024_M87_PerShad}, and 2021\cite{EHT_2025_obs2021}. In 2017, EHT observations of \mes\ were conducted with 7 separate facilities\cite{EHT_2019_Array}: the Atacama Large Millimeter/submillimeter Array (ALMA)\cite{Wootten_2009_ALMA, Goddi_2019_ALMA}, the Atacama Pathfinder Experiment telescope (APEX)\cite{Gusten_2006_APEX}, the Large Millimeter Telescope (LMT)\cite{Hughes_2010_LMT}, the Pico Veleta IRAM 30 m telescope (PV)\cite{Greve_1995_PV}, the Submillimeter Telescope Observatory (SMT)\cite{Baars_1999_SMT}, the James Clerk Maxwell Telescope (JCMT)\cite{Hills_1985_JCMT}, and the Submillimeter Array (SMA)\cite{Ho_2004_SMA}, which heterogeneously and sparsely sample the Fourier plane as the Earth rotates. Additionally, the Greenland Telescope (GLT)\cite{Inoue_2014_GLT, Chen_2023_GLT} participated in the 2018 observations for the first time as part of the EHT array, which improved the aperture coverage, particularly along the North-South direction. For 2021, the 12m Kitt Peak (KP)\cite{Freund_2014_KP12m} Telescope and the upgraded Northern Extended Millimeter Array (NOEMA)\cite{Kim_2023_NOEMA} facility joined EHT observations for the first time.

The 2017 observations of \mes\ are spread over four separate epochs (April 5, 6, 10, 11), while observations in 2018 and 2021 are separated into two epochs each (April 21/25 for 2018, April 13/18 for 2021). The datasets with the best single-day aperture coverage each year are April 11, April 21, and April 18, which we adopt as our fiducial datasets, while datasets on other days are used as cross-validation. We consider the data taken over two $\sim2$-GHz bands centered at 227.1 (low) and 229.1 GHz (high), which we merge into a single dataset.

The interferometric data on \mes\ are comprised of complex visibilities, correlated from signals between pairs of stations, referred to as a baseline, which are, by the van Cittert-Zernike theorem\cite{Zernike_1938}, related to the source brightness distribution by a two-dimensional Fourier transform after certain approximations\cite{TMS}. Visibilities can be sequenced into scans, based on the time difference between observations, from which we obtain coherently averaged visibilities. In this work, closure invariants are generally computed from scan-averaged visibilities. With typical scan lengths on the order of five minutes, temporal averaging can potentially introduce residuals from time-dependent gain fluctuations to closure invariants. We have internally validated the robustness of our imaging products by computing closure invariants at the smallest possible cadence of the science-ready dataset (typically 10s) prior to scan-level averaging. Following the procedure in the publicly available \mes\ imaging pipeline\footnote{\href{https://github.com/eventhorizontelescope/2019-D01-02}{https://github.com/eventhorizontelescope/2019-D01-02}}, we rescale short baselines to the total flux density estimated from the ALMA and APEX baseline in order to excise the contribution from extended flux. We have also internally validated that the short baseline rescaling procedure does not affect the recovered morphology from our imaging algorithm. For quantitative comparison, we generate fiducial images from the publicly available \ehtim\ pipeline to represent EHTC reconstructions.

\subsection*{Closure invariants} 
Closure invariants are calibration-independent interferometric measurables produced from a generalised formalism of traditional closure phase\cite{Jennison_1958,Thyagarajan_2022_CPhase} and amplitudes\cite{Twiss_1960}, with both a co-polar version\cite{Thyagarajan_2022_CI} and polarimetric version\cite{Samuel_2022}. \gendirect\ utilises co-polar closure invariants constructed from elementary triangular plaquette variables called ``advariants''. The complete minimal set of independent closed triads can be constructed from an array of $n$ stations by selecting all pairs of array elements $(a,b)$ pinned on a fixed reference vertex indexed at 0,
\begin{equation}
    \mathcal{A}'_{\rm{0ab}} = \mathcal{V}'_{0a}(\mathcal{V}'^{*}_{ab})^{-1}\mathcal{V}'_{b0} = \abs{g_0}^2\mathcal{A}_{\rm{0ab}}\,,
\end{equation}
where $\mathcal{V}'_{ab}$ are corrupted visibilities measured from the baseline between stations $a$ and $b$, and $\abs{g_0}^2$ is the unknown scaling factor associated with the reference station which is shared by all complex advariants constructed in this way. By normalising all advariants by the $L_2-$norm, for example, we produce $N_\textrm{s}^2 - 3N_\textrm{s} + 1$ real-valued independent closure invariants, whose form is different but carry information equivalent to that in traditional closure amplitudes and phases. In the co-polar formalism of closure invariants, traditional closure phases can be identified trivially from the phases of the advariants, while closure amplitudes can be derived from specific pairs of advariants. A key advantage of this formalism over traditional closure amplitudes and phases is that it avoids the heterogeneous treatment that phases and amplitudes necessitate. In internal validation tests, closure invariants as the sole likelihood term was found to be preferable over the independently-weighted (log-)closure amplitudes and closure phases in terms of convergence and data efficiency.

Various selections of the reference station can affect covariances between closure terms, despite carrying equivalent information. A different selection of reference station can influence the signal-to-noise ratio in the derived closure invariants, but does not significantly influence the resulting image reconstruction\cite{Lai_2026_3C279CenA}. For this work, in order to avoid bias from a single reference station selection, we present the combined results from two \gendirect\ models for each dataset, trained consistently on the first and the second-most sensitive stations as the reference stations in each scan. Selecting from among the most sensitive stations maximises closure invariant SNR and minimises covariances between closure terms\cite{Blackburn_2020}.

\subsection*{\gendirect\ imaging} \label{methods:gendirect}
The Generative Deep learning Image Reconstruction with Closure Terms algorithm (\gendirect)\cite{Lai_2025_GenDIReCT} is an imaging algorithm which utilises co-polar closure invariants\cite{Thyagarajan_2022_CI} for calibration-independent imaging under the constraints of sparse aperture coverage characteristic of VLBI data. The architecture is comprised of a latent diffusion model\cite{Rombach_2021_StableDiff} conditioned on closure invariants and a multi-layer convolutional neural network (CNN) used to optimise the closure invariants likelihood. Traditional imaging algorithms, such as hybrid mapping and self-calibration followed by deconvolution, begin with a simple initial model and iteratively solve for antenna-based complex gains and the image model to converge upon a final solution that matches the measured data. 
These methods can be sensitive to and biased by image initialisation and calibration errors, particularly in sparse and low signal-to-noise measurements\cite{Readhead_1978, Pearson_1984_SelfCal, Carilli_2022}. Our algorithm bypasses the limitations of conventional techniques by utilising closure invariants as the robust interferometric measurable and produces an image reconstruction directly in a single forward pass. Additionally, in contrast to the newer regularised maximum likelihood methods which can also image using closure invariants as the sole data term\cite{Akiyama_2017_SMILI, Chael_2018_ehtim, Mueller_2022_doghit, Muller_2023_MOEAD}, our algorithm provides a generative solution that does not require a user-specified, hand-tuned image initialisation at inference time. Other machine-learning interferometric imaging techniques have been presented in the literature, based on CNNs\cite{Sureau_2020, Nammour_2022, Schmidt_2022, Chiche_2023, Terris_2023}, normalising flow\cite{Sun_2022_adeep, Feng_2024}, denoising diffusion\cite{Drozdova_2024, Feng_2024}, classifiers\cite{Rustige_2023, Thyagarajan_2024_Lucas}, adversarial networks\cite{Geyer_2023, Rustige_2023}, and transformers\cite{Lai_2025_DIReCT}. While there are some similarities with other diffusion-based imaging algorithms\cite{Drozdova_2024, Feng_2024}, \gendirect\ is unique in its closure-conditioned latent diffusion model, and importantly, in its use of homogeneous closure invariants rather than the traditional closure phases and closure amplitudes that require heterogeneous treatment. In this section, we provide a brief overview of our algorithm and its past applications.

The neural network architecture is composed of two distinct components. The first component is a conditional UNet\cite{Ronneberger_2015_Unet} based on textually-conditioned image generation architectures\cite{Rombach_2021_StableDiff} with denoising diffusion probabilistic models\cite{DDPM_ho_2020}, which has emerged as a highly effective image generation technique, capable of modelling complex data distributions. Our implementation is a latent diffusion model consisting of three convolutional layers set in a symmetric contracting and expanding pathways with residual connections\cite{He_2016_RESNET} from the contracting path that map features to their mirrored expanding layers. In the contracting path, we use one down-sampling layer, followed by two down-sampling layers with cross attention, where the closure invariant features adjust convolutional features via the cross-attention mechanism, which conditions the denoising process. On the up-sampling path, we implement two up-sampling layers with self-attention and one final up-sampling layer without attention. By having no cross-attention layers in the up-sampling path, the result is a robust conditional denoising network that performs well even with noise from real observations. The latent shape is 4 channels of $16\times16$ grids, which is down-sampled from the original $64\times64$ total intensity image grid using a pretrained autoencoder\cite{Lai_2025_DIReCT}. The second component of the \gendirect\ architecture is a standard multi-layer CNN composed of three sequential down-sampling layers, designed to compress a large set of diffusion-sampled images (each considered a separate channel) into a representative data-consistent single-channel image reconstruction. We note that we have experimented with multiple design variations and found this version to be suitable in all tested cases, but better-performing alternatives for specific datasets may be possible.

\gendirect\ works by learning realistic image characteristics (such as positivity and smooth brightness distribution) from a non-astronomical image training dataset, primarily composed of augmented CIFAR-10 natural images\cite{Krizhevsky09_CIFAR10}. The conditional latent diffusion model generates sensible image priors which the multi-layer CNN uses as input to optimise the data likelihood and refine the final image reconstruction. As a generative imaging algorithm, our algorithm has the potential to handle degeneracies and visualise multimodalities in the solution space, although they should not be interpreted as image reconstruction uncertainty under a Bayesian framework. Generally, appearance of minor artifacts in low-sensitivity regions, different levels of smoothness and sharpness, or super-resolution variations are acceptable consequences of a diversity in potential solutions to the ill-posed image reconstruction problem. However, methodological and hyperparameter sensitivity, or inability to satisfy the underlying measurements constitute signs that a reconstruction is not robust. By training on unconvolved ground-truth models, \gendirect\ achieves super-resolved reconstructions and enables generative visualisation of structural variations that remain strictly consistent with the observed data. Accordingly, unless otherwise indicated, we present unconvolved \gendirect\ outputs optimised to be consistent with the measurements.

We have extensively tested our algorithm on synthetic datasets with the EHT or next-generation EHT (ngEHT) array\cite{Lai_2025_GenDIReCT} as well as non-horizon-scale EHT datasets on active galactic nuclei, \cagn\ and \cena\cite{Lai_2026_3C279CenA}. On synthetic datasets, our algorithm generalises well to untrained morphologies and maintains $>90\%$ image similarity with ground truth images at signal-to-noise levels well below the datasets of \mes. Moreover, its performance on quantitative metrics is comparable to other state-of-the-art imaging algorithms on the ngEHT total intensity analysis challenge\cite{Roelofs2023_ngeht-challenge}. For the EHT dataset on \cagn, our reconstruction is 98\% similar to the EHTC reference\cite{Kim_2020_3C279}, reproducing the core-jet morphology across multiple independent epochs and enabling direct measurement of the superluminal relative proper motion between components in the image-domain. For the EHT dataset on \cena, the EHTC reference\cite{Janssen_2021_CenA} is reproduced with 91\% similarity with statistically compatible brightness ratios and opening angle between the North and South edge-brightened jet ridges. These extensive tests on real and synthetic data demonstrate that our algorithm is capable of consistently recovering diverse morphological structures, without manual intervention or prescriptive priors. 

Prior to applying to the real data, each trained model is extensively tested on an array of internal validation tests, which include quantifying its performance degradation with out-of-distribution image morphologies and varying levels of data corruption, accounting for the heterogeneous sensitivity of the array\cite{Lai_2026_3C279CenA}. Only after passing internal validation are the models unblinded on real data to minimise risk of confirmation bias. No alterations are made to the model between the validation procedure and the unblinding, or post-unblinding. 

\begin{figure}[htb]
\begin{center}
\resizebox{16cm}{!}{\includegraphics[trim={0 0 0 0},clip]{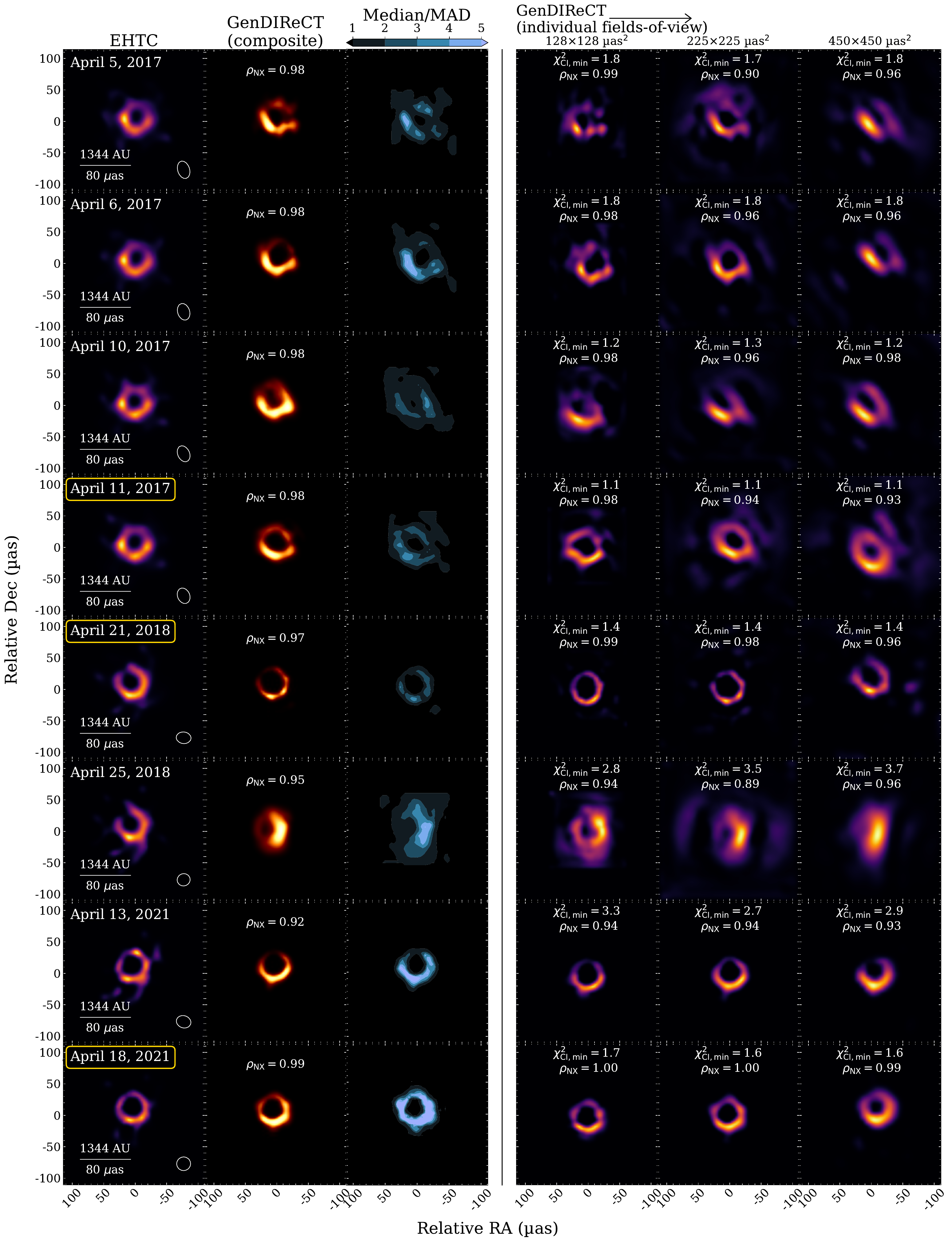}}
\end{center}
\caption{(Left) Fiducial reconstructions from \ehtim\ for the single-epoch datasets. The highest quality dataset of each year is highlighted with a gold border. (Right panels) Corresponding composite \gendirect\ reconstruction, median to median absolute deviation ratio, and median representative reconstructions across multiple fields of view. Data fidelity ($\chi^2_{\rm CI, min}$) on median reconstructions and image similarity ($\rho_{\rm NX}$) scores are illustrated, where $\rho_{\rm NX}$ is measured relative to the EHTC fiducial image reconstruction. All panels are displayed on a $225\times225\,\mu{\rm as}^2$ window.
}
\label{fig:median-results}
\end{figure}

For this work, we train our machine learning architecture with synthetic data derived using the data coverage from all four 2017 and both 2018/2021 observing days independently, combining the two 2-GHz bands into a single scan-averaged dataset for each epoch. We choose $128\times128\,\mu\rm{as}^2$, $225\times225\,\mu\rm{as}^2$, and $450\times450\,\mu\rm{as}^2$ reconstruction fields, each on $64\times64$ pixel grids as pre-determined by the architecture. Generally, we represent the aggregate reconstruction for each dataset by the median of all aligned reconstructions. We find the median representation to be more robust against outliers or source alignment issues, and a better representation of the bulk of individual image reconstructions. The aggregate reconstructions for all datasets and fields of view are shown in Extended Data Figure \ref{fig:median-results} alongside EHTC comparison reconstructions, \gendirect\ composites, and a median to median absolute deviation (MAD) ratio panel. The median-to-MAD ratio is presented as a visualisation of \gendirect's reconstruction confidence\cite{Lai_2025_GenDIReCT}, where features of low variance relative to the median pixel intensity (i.e. high reconstruction confidence) are highlighted with filled contours. 

\begin{figure}[htb]
\begin{center}
\resizebox{16cm}{!}{\includegraphics[trim={0 0 0 0},clip]{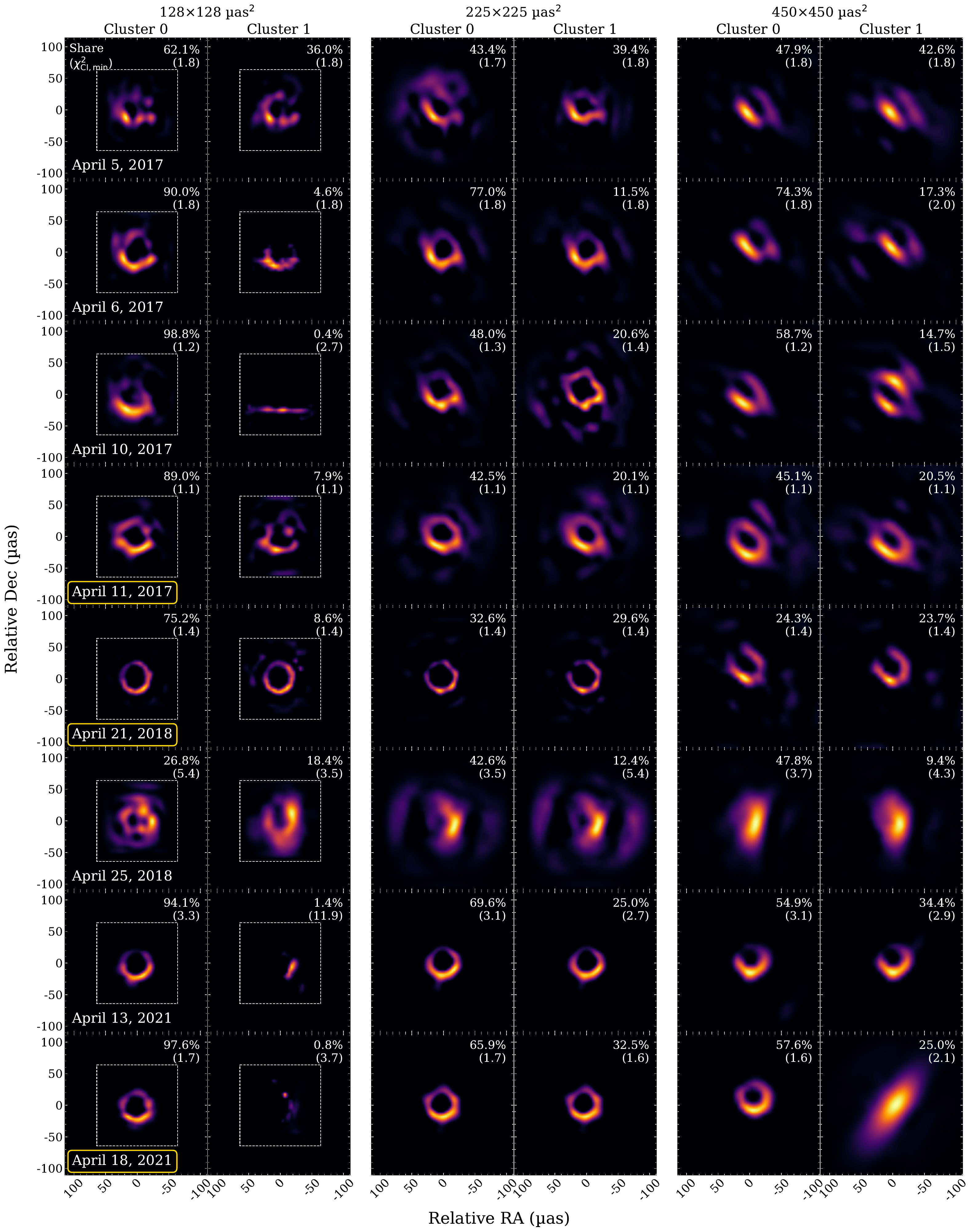}}
\end{center}
\caption{Clusters of \gendirect\ reconstructions across $128\times128$, $225\times225$, and $450\times450\,\mu$as$^{2}$ fields of view. The two most-populated clusters for each field of view are presented in descending order from left to right. Each row represents a separate single-epoch dataset, and the highest quality dataset of each year is highlighted with a gold border. The population of each cluster is represented by a percentage fraction alongside the minimum $\chi^2_{\rm CI}$ in parentheses. Displays containing the five most-populated clusters can be found in the \gendirect\ repository.
}
\label{fig:M87_multiFoV_clusters}
\end{figure}

One of the advantages of the generative imaging process of our algorithm is the potential to explore different imaging solution modes. We consider all observation epochs and all fields of view, presenting the median reconstructions from the most populated solution clusters in Extended Data Figure \ref{fig:M87_multiFoV_clusters}. Image reconstructions are clustered by Hamming distance in their perceptual hash representation\cite{Lai_2026_3C279CenA} to illustrate multimodal solutions. Each cluster is also presented with the best data fidelity score in the parentheses. Often, more than 50\% of all reconstructions reside within the top two most populous and perceptually similar clusters. This clustering behavior justifies the pixel-wise median as a statistically sound and accurate aggregate representation. The \mes\ clustering results clearly demonstrate that the differences between reconstruction clusters for the observation days with the highest quality data are in the surface brightness and extent of extraneous amorphous emission surrounding the ring, rather than fundamental changes to the ring structure itself. For other datasets, solution clusters which are not as clearly ring-like are typically 
of sparse occurrence
in comparison to the bulk of reconstructions and exhibit a lower data fidelity score. On all fields of view, there is no non-ring solution which is favoured over ring-like morphology in the share of the reconstructed image population or data fidelity score. The most ambiguous case is that of the April 25, 2018 dataset, which is missing data from the LMT, resulting in reduced image fidelity compared to the April 21, 2018 observations. The absence of LMT-related baselines created a substantial gap in $uv$-coverage at spatial frequencies corresponding to $\sim1\,{\rm G\lambda}$, as the LMT–SMT baseline provided the unique long-baseline constraint for structure in this regime\cite{EHT_2024_M87_PerShad}.

\begin{figure}[htb]
\begin{center}
\resizebox{16cm}{!}{\includegraphics[trim={0 0 0 0},clip]{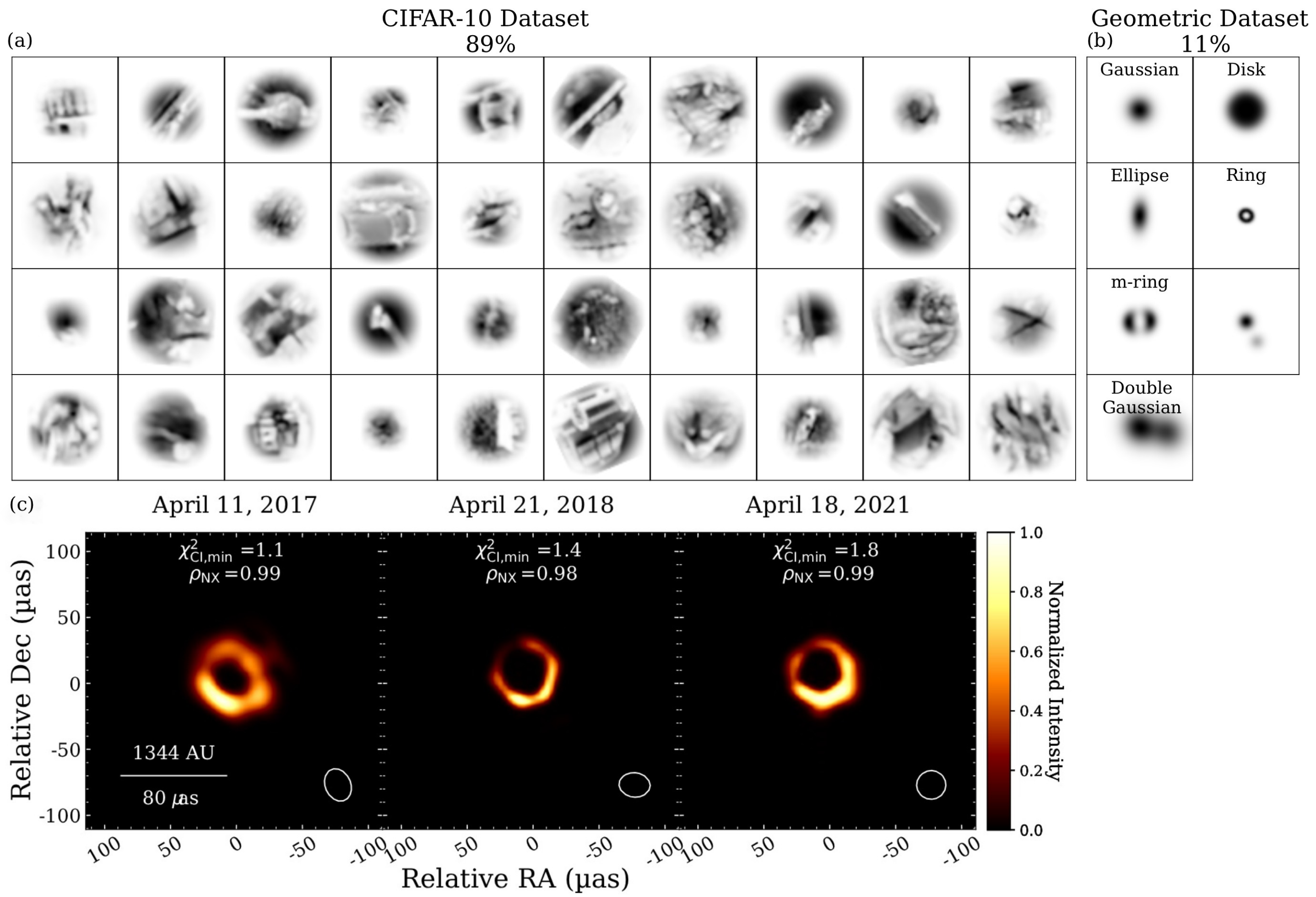}}
\end{center}
\caption{Example grid of augmented image morphologies used for training \gendirect\ and reconstructions in the absence of the geometric dataset. (a) Image grid shows an assortment of augmented non-astronomical CIFAR-10 images which constitute the bulk of the training dataset. (b) Panels illustrate classes of simple geometric shapes used in training, which are not required to produce robust image reconstructions of \mes. Percentages indicate the fraction of the total training dataset composed by each class. (c) Alternative \gendirect\ composites for the highest quality single-day datasets from observations in 2017, 2018, and 2021, without geometric shapes in the training data. The image similarity score is with reference to the EHTC fiducial reconstructions produced with \ehtim.}
\label{fig:training-data-and-recon}
\end{figure}

Unless otherwise stated, all models described in this work are trained on augmented non-astronomical images from the CIFAR-10 dataset\cite{Krizhevsky09_CIFAR10}, as well as simple geometric shapes, such as Gaussians, disks, rings, ellipses, and $m$-rings\cite{Roelofs_2023_mring} of first and second order\cite{Lai_2025_GenDIReCT}. We present an example grid of augmented training images in Extended Data Figure \ref{fig:training-data-and-recon}. Once trained, we run the imaging sequence for 96 total GPU-hours on a single NVIDIA-H100 for each dataset, producing several hundred ($\sim300$) individual image reconstructions, each drawing from 4096 diffusion samples. The exact number of final image reconstructions depends on data volume.
Although our algorithm has been shown to be generalisable to arbitrary source morphologies\cite{Lai_2025_GenDIReCT}, there is a concern that the presence of crescents and rings in the training dataset may be a source of bias for the image reconstruction. Therefore, we also investigate the outcome of training our architecture without geometric shapes in the training dataset and we find no significant difference in the image reconstruction (Extended Data Figure \ref{fig:training-data-and-recon}).

\subsection*{Metrics} 
We employ the reduced $\chi^2_{\rm{CI}}$ metric to evaluate the consistency between our image reconstructions with the 
measurements. 
The reduced $\chi^2_{\rm{CI}}$ metric is defined as the sum of error-normalised residuals, divided by the number of independent data terms, 
\begin{equation}
        \chi^2_{\rm{CI}} = \frac{1}{N_{\rm{ci}}}\sum_{i}^{N_{\rm ci}}\left[\frac{\left(\mathcal{C}(A)_i - C_i\right)^2}{\sigma_i^2}\right]\,, \label{eq:cnn-loss}
\end{equation}
where $\mathcal{C}$ is an operator which maps the reconstructed image to its closure invariants, indexed by $i$, for a fixed observation arrangement, and $C_i$ is the corresponding closure invariant of the measurement. The total number of closure invariants is denoted by $N_{\rm ci}$ and $\sigma_i^2$ is the estimated uncertainty on the $i$-th closure invariant. We also define the normalised residual of any data point, $i$, as $\left(\mathcal{C}(A)_i - C_i\right)/\sigma_i$. 

\begin{figure}[htb]
\begin{center}
\resizebox{16.5cm}{!}{\includegraphics[trim={0 0 0 0},clip]{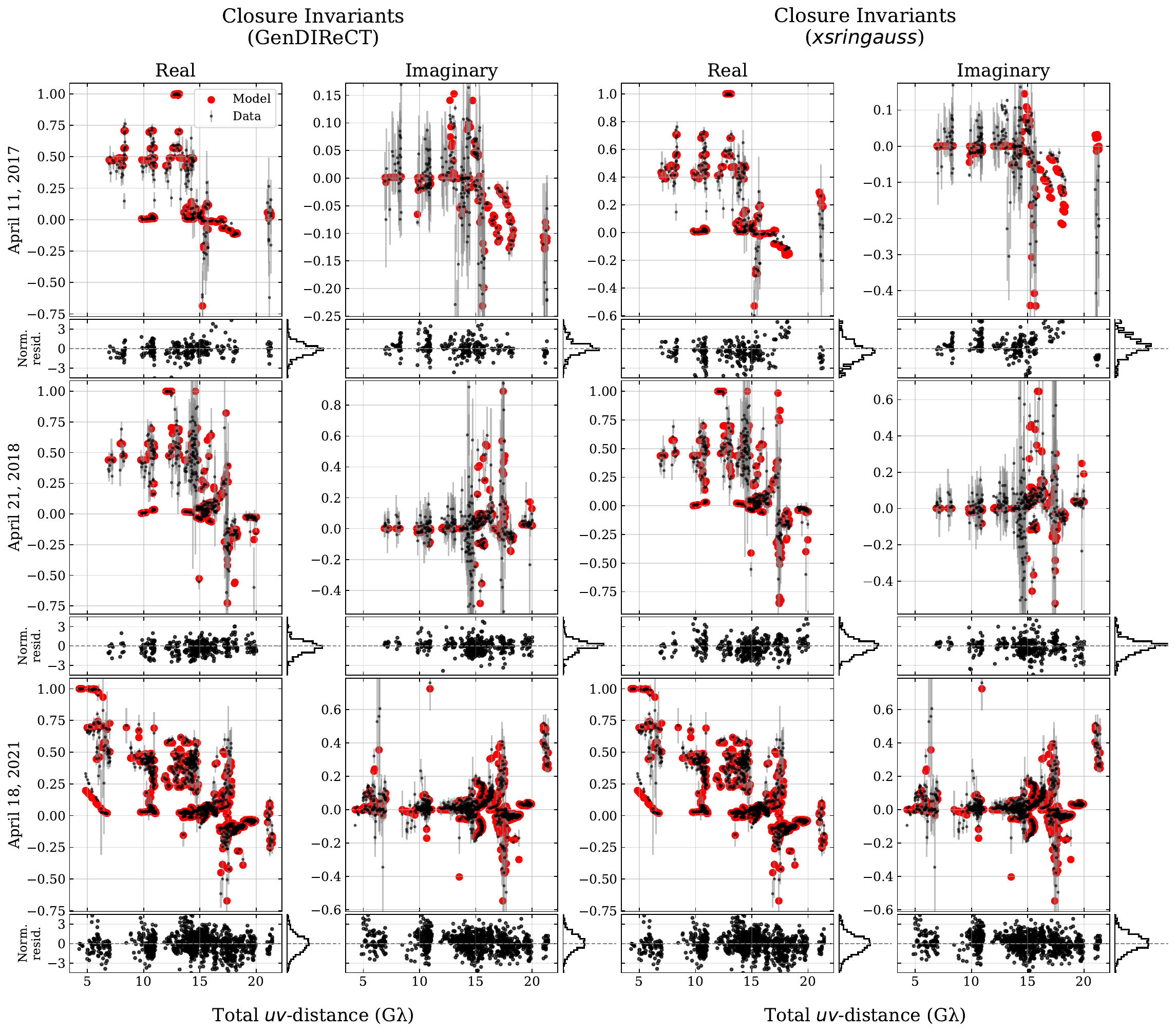}}
\end{center}
\caption{\textbf{Data fidelity from the best reconstructions on each dataset}, as determined by the minimum $\chi^2_{\rm ci}$ of all \gendirect\ (left) and \xsrg\ geometric modelling (right) reconstructions for each respective dataset, where closure invariants are presented as real and imaginary values across the summed $uv$-distance in G$\lambda$ between all three participating stations. The residuals normalised by the expected uncertainty are presented on separate adjoining panels.}
\label{fig:data-adherence}
\end{figure}

While propagating the estimated uncertainty from each radio station's system equivalent flux density (SEFD) is straightforward, the total error on each closure invariant is affected by other sources of systematic errors, including time-dependent errors from long coherent visibility averaging, and limitations of the linear error propagation model. Generally, the fractional systematic uncertainty budget of 3C~279 and M87 data is $\sim5\%$, which is similar in magnitude as measured inter-pipeline differences\cite{EHT_2019_Data}. However, in this work, we present our results without additional systematic error tolerance and we consistently use as reference the highest quality single-day \hops-reduced pre-scan-averaged dataset of each observation year (April 11 for 2017, April 21 for 2018, April 18 for 2021) when evaluating the reduced $\chi^2_{\rm{CI}}$ to ensure a consistent metric across reconstructions. For these datasets, we illustrate the data fidelity from the best \gendirect\ reconstruction and the \xsrg\ geometric model in Extended Data Figure \ref{fig:data-adherence}. A near-unity reduced $\chi^2_{\rm{CI}}$ value confirms that a reconstruction is self-consistent with the measured closure invariants, but does not guarantee that the recovered morphology is correct in an absolute sense, making the reduced $\chi^2_{\rm{CI}}$ a necessary but insufficient metric for an accurate reconstruction.

In some cases, it is useful to measure the similarity between reconstructions in the image domain, especially when comparing the \gendirect\ reconstructions to the results from the EHTC imaging pipelines. We measure the image-to-image correspondence with the maximum normalised cross-correlation metric, $\rho_{\rm{NX}}$, which has the advantage of being independent of image translation and total flux scaling. The metric has been widely used to quantitatively measure image similarity, particularly with image reconstructions produced using only closure data terms\cite{Chael_2018_ehtim, Roelofs2023_ngeht-challenge, Lai_2025_GenDIReCT, Lai_2026_3C279CenA}. The $\rho_{\rm{NX}}$ metric is defined as,
\begin{equation}
    \rho_{\rm{NX}}(I_1, I_2) = \frac{1}{M}\max\abs{{\mathcal{F}^{-1}\{\mathcal{F}\{\hat{I_1}\}\mathcal{F}\{\hat{I_2}\}^*\}}}\,,
\end{equation}
where, $I_1$ and $I_2$ are two images being compared, and the hat operator, $\hat{I} = (I - \bar{I})/\sigma_I$, normalises each image $I$ by its respective flux mean and standard deviation. We denote the forward and inverse Fourier transform as $\mathcal{F}$ and $\mathcal{F}^{-1}$, respectively. The maximum operator ensures that the metric is independent of image translation and it is also useful for aligning separate image reconstructions. To minimise the negative influence of high-frequency artifacts on the maximum normalised cross-correlation, we convolve each image by the fitted clean beam prior to computing $\rho_{\rm{NX}}$. Throughout this work, we use $\rho_{\rm{NX}}$ to evaluate the inter-method consistency in resulting image reconstructions, often between the \gendirect\ and EHTC fiducial imaging pipelines or simulated images. As there is no ground truth reconstruction for \mes, the $\rho_{\rm{NX}}$ metric does not constitute an independent ground truth validation metric, but as a image-domain similarity metric between various imaging algorithms.

\subsection*{Image-domain feature extraction} 
The image-domain feature extraction method presented here follows the framework established by the EHTC in their initial analysis of \mes\cite{EHT_2019_Imaging}. We outline the reproduction of this algorithm below. Given any plausibly ring-like image, we first search for the ring center, $(x_0, y_0)$, by evaluating all candidate centroid positions for the one which minimises the normalised radial peak dispersion,
\begin{equation}
    (x_0, y_0) = {\rm argmin}\left[\frac{\sigma_r(x, y)}{\bar{r}_{\rm pk}(x, y)}\right]_{(x, y)}\,,
\end{equation}
where $\bar{r}_{\rm pk}(x, y)$ is the radius of peak brightness averaged over all angles, as
\begin{equation}
    \bar{r}_{\rm pk}(x, y) = \left<{\rm argmax}_r [I(r, \theta; x, y)] \right>_{\theta \in [0, 2\pi]}\,,
\end{equation}
where $I(r,\theta; x,y)$ is the intensity distribution of the image in radial coordinates, centered on $(x, y)$. The dispersion, $\sigma_r(x, y)$, is the corresponding standard deviation of ${r}_{\rm pk}(\theta ;x, y)$ across all angles, $\theta$. The ring diameter and its uncertainty can then be identified from $\bar{r}_{\rm pk}(x, y)$ and  $\sigma_r(x, y)$, respectively, given the measured ring center, $(x_0, y_0)$. 

Once the ring center is determined, most other characteristic features can be extracted directly from the polar image, including: the ring width $w_r$, orientation $\eta$, azimuthal asymmetry $A$, and central brightness fraction $f_c$. The ring width, $w_r$, is determined by measuring the mean full-width at half-maximum (FWHM) at each radial slice for the given ring center,
\begin{equation}
    w_r = \left< {\rm FWHM} \left[I(r,\theta) - I_{\rm floor} \right]\right>_{\theta}\,,
\end{equation}
where the floor value $I_{\rm floor} = \left< I(r=2r_{\rm pk}, \theta) \right>_{\theta}$ is used to correct for any non-zero residual flux outside the ring. The uncertainty on the ring width, $\sigma_w$, is measured by the standard deviation in FWHM across all $\theta$. 

The orientation angle, $\eta$, of the ring (measured East of North) is the circular mean of the first angular mode from $r_{\rm in} = r_{\rm pk} - w_r/2$ to $r_{\rm out} = r_{\rm pk} + w_r/2$, as
\begin{equation}
    \eta = \left< {\rm Arg}\left[ \int_0^{2\pi} I(\theta)e^{i\theta}d\theta \right]\right>_{r \in [r_{\rm in}, r_{\rm out}]}\,,
\end{equation}
and the corresponding uncertainty, $\theta_\eta$, is the standard deviation of the measured angles for $r \in [r_{\rm in}, r_{\rm out}]$.

The azimuthal asymmetry, $A$, is the mean of the normalised amplitude in the first angular mode from $r_{\rm in}$ to $r_{\rm out}$, which is,
\begin{equation}
    A = \left<\frac{\abs{\int_0^{2\pi} I(\theta)e^{i\theta}d\theta}}{\int_0^{2\pi} I(\theta)e^{i\theta}d\theta}\right>_{r \in [r_{\rm in}, r_{\rm out}]}\,,
\end{equation}
where $\sigma_A$ is the standard deviation in the measured asymmetry over $r \in [r_{\rm in}, r_{\rm out}]$. Perfect azimuthal symmetry corresponds to a value of $A = 0$, whereas a delta function with all flux concentrated at a single orientation angle corresponds to a value of $A = 1$. 

\begin{figure}[htb]
\begin{center}
\resizebox{14cm}{!}{\includegraphics[trim={0 0 0 0},clip]{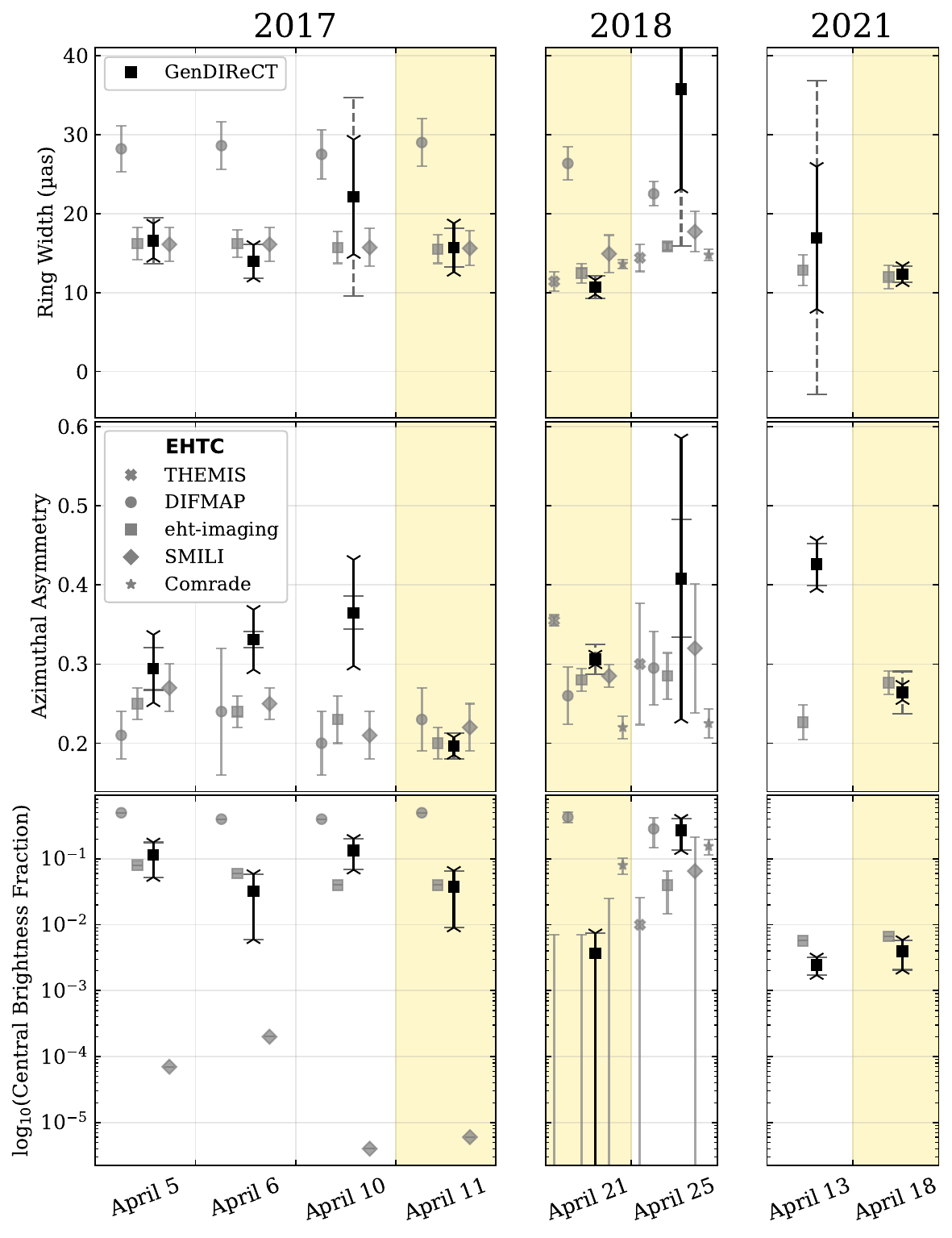}}
\end{center}
\caption{Similar to Figure \ref{fig:ringdiameter}, but for measured ring width, azimuthal asymmetry and central brightness fraction.}
\label{fig:ring-params-ext}
\end{figure}

Finally, the central brightness fraction, $f_c$, is the ratio between the mean interior depression brightness to the mean ring surface brightness, as,
\begin{equation}
    f_c = \frac{\left<I(r, \theta)\right>_{\theta, \,r\in[0, \,0.2 r_{\rm pk}]}}{\left<I(r_{\rm pk}, \theta)\right>_\theta}\,,
\end{equation}
where the interior brightness is measured from the average flux in a small disk around the ring center. Though we do not define an uncertainty for a single measurement of $f_c$, we derive its uncertainty from multiple measurements across the three different fields of view used in \gendirect\ reconstruction. 
Figure \ref{fig:ringdiameter} in the main text presents the measured ring diameter and orientation across all datasets considered in this work. We present the measurements for ring width, azimuthal asymmetry, and central brightness fraction in Extended Data Figure \ref{fig:ring-params-ext}.

The EHTC derived ring parameters using the \rex\ module implemented as part of \ehtim. Ring properties from three different imaging algorithms (\difmap\cite{Shepherd_2011_DIFMAP}, \ehtim\cite{Chael_2018_ehtim}, \smili\cite{Akiyama_2017_SMILI}) were reported for the 2017 data\cite{EHT_2019_Imaging}, two Bayesian imaging algorithms (\themis\cite{Broderick_2020_Themis}, \comrade\cite{Tiede_2022_Comrade, Tiede_2026_Comrade}) were included for the 2018 datasets\cite{EHT_2024_M87_PerShad}, and for 2021 data, one additional \clean-based algorithm (\textsc{GPCAL}\cite{Park_2021_GPCal}) and two additional regularised maximum likelihood algorithms (\doghit\cite{Mueller_2022_doghit}, \textsc{MOEA/D}\cite{Muller_2023_MOEAD}) were added. However, for the 2021 datasets, the EHTC adopted an alternative template-matching image-domain feature extraction algorithm\cite{EHT_2025_obs2021}. As such, to maintain methodological consistency, we have reproduced a single \ehtim\ reconstruction to use as the EHTC reference.

When comparing ring diameter between datasets, image reconstruction algorithms, and across multi-year epochs, there is potential for systematic bias arising from the finite resolution effect, where variations in ring width can shift the apparent diameter derived from a direct image-domain feature analysis.
Recent EHTC analyses employ a forward-modelling procedure to mitigate diameter-width degeneracy\cite{EHT_2025_obs2021}. In this work, we choose to maintain methodological consistency across all reconstructions by deriving diameters directly from the image-domain without further adjustment. Therefore, the apparent variation observed between imaging algorithms and individual datasets may be influenced by each imaging method's finite resolution. Despite this, the apparent ring diameter derived from our reconstructions remains self-consistent across all independent datasets.

\subsection*{Geometric modelling} 
Our complementary Bayesian analysis utilises the geometric \xsrg\ model\cite{Benkevitch_2016_sdisks, EHT_2019_ShadowMass}, which was designed to simulate the appearance of quasi-Kerr black hole accretion disk radio images\cite{Broderick_2014_quasiKerrImages}. The advantage of geometric modelling is that visibilities can be derived analytically from simple algebraic expressions and fit to the observed measurements, which significantly reduces the computational requirement of modelling compared to complex and highly non-linear physical models, particularly when applied with Bayesian sampling algorithms. Moreover, as long as the object is plausibly described by the model, Bayesian sampling methods applied with geometric models enable direct inference of key features of interest from the observed object with associated uncertainties. In this section, we provide a brief description of the \xsrg\ model, the additional components we included to form the model for fitting, and the complete analytical visibility function. We develop the \textsc{BASIS}\cite{BASIS_Zenodo} software package for Bayesian analysis with the \xsrg\ model independent from existing software from the EHTC. For the following discussion, we use $\mathcal{F}$ to denote the forward Fourier transform operator.

The \xsrg\ model is parameterised with the following variables\cite{EHT_2019_ShadowMass}. The outer disk radius, $R_{\rm out}$, and inner disk radius, $R_{\rm in}$, create the centralised ``top-hat crescent'', where the offset of the inner disk centroid is controlled by the eccentricity, $\epsilon \in [0, 1]$. At $\epsilon = 0$, the centers of both disks are co-located at the origin, and at the maximum eccentricity, $\epsilon = 1$, the edge of the inner disk meets the edge of the outer disk. The fading parameter, $f \in [0, 1]$, controls the surface brightness gradient across the crescent, with the maximum brightness located at the tip of the crescent, where the crescent width is maximum. When $f=0$, the crescent surface brightness is uniform and when $f=1$, the surface brightness fades to zero at $x = -R_{\rm out}$. Note that we have deliberately adjusted the definition of fading from its original conception\cite{Benkevitch_2016_sdisks}. 

The \xsrg\ model also includes an elliptical Gaussian centered on the inner disk edge where the crescent is widest. Its flux is determined by a fraction, $f_g$, of the total flux and its width is set by $\sigma_e$, with axial ratio, $a_g$. The longest axis is always oriented perpendicular to the crescent's axis of symmetry. A circular Gaussian with flux, $V_f$, and width, $\sigma_f$, centered at $x_0$ represents the floor component internal to the crescent. To complete the base \xsrg\ model, we blur the image with a circular Gaussian kernel of width, $\sigma_b$, and rotate the entire image by angle, $\phi$.

Additional components include a larger-scale circular background Gaussian of flux, $V_{\rm bg}$, and width, $\sigma_{\rm bg}$, centered at the origin, which is added to include short-spacing flux that is not probed by the majority of the long-baseline array. Finally, two additional nuisance elliptical Gaussians are included, each with independent parameters for flux, width, axial ratio, position angle, angular offsets, and Gaussian blur, to capture extraneous amorphous emission that is not modelled by the ring alone and as additional degrees of freedom to absorb the effects from measurement errors and systematic uncertainties.

In the Fourier domain, the composite model is a linear combination of elementary components with analytic Fourier transforms. The general Fourier transform (represented by tilde) of a disk with radius, $R$, and amplitude, $V$, is given by,
\begin{equation}
    \tilde{D}(\rho) = \frac{VR}{\rho} J_1(2\pi\rho R)\,,
\end{equation}
where $\rho$ is the Fourier-domain radial coordinate and $J_1$ is a first-order Bessel function of the first kind. The Fourier transform of a Gaussian with flux, $V$, and width, $\sigma$, is,
\begin{equation}
    \tilde{G}(\rho) = V e^{-2\pi^2\sigma^2\rho^2}\,,
\end{equation}
where the Fourier transform of asymmetric Gaussians are determined by exploiting the similarity property of the Fourier transform, i.e. $\mathcal{F}\left(G(ax, by)\right) = \frac{1}{\abs{ab}}\tilde G(\frac{u}{a}, \frac{v}{b})$. Multiplication in the image domain of the crescent by the sloped plane is represented in the Fourier domain by taking a derivative, 
\begin{equation}
    \mathcal{F}(xF(x)) = \frac{i}{2\pi}\frac{d}{ds}\tilde{F}(s)\,,
\end{equation}
where for our specific linear slash function, $l(x)$, the corresponding operation in the Fourier domain is,
\begin{equation}
    \mathcal{F}\left(l(x)F(x)\right) = h\left(1-\frac{f}{2} + \frac{i}{2\pi R_{\rm out}} \frac{f}{2} \frac{d}{ds} \right)\tilde F(s)\,,
\end{equation}
where the derivatives of Bessel functions of the first kind, $J_n$, can be expressed in terms of other Bessel functions, $J_{n\pm1}$, as,
\begin{equation}
    \frac{d}{d\rho}\,J_1(2\pi\rho R) = \pi R\left[J_0(2\pi\rho R) - J_2(2\pi\rho R)\right]\,,
\end{equation}
and $h$ is responsible for normalising the source to the desired total flux, $V$, which is dependent on the source structure. For the \xsrg\ model, the normalisation factor is,
\begin{equation}
     h = \frac{2V}{\pi\left[R_{\rm out}^2 (2-f) + R_{\rm in}^2\left(f(1+\frac{x_0}{R_{\rm out}}) - 2\right)\right]}\,.
\end{equation}

\begin{figure}[htb]
\begin{center}
\resizebox{16cm}{!}{\includegraphics[trim={0 0 0 0},clip]{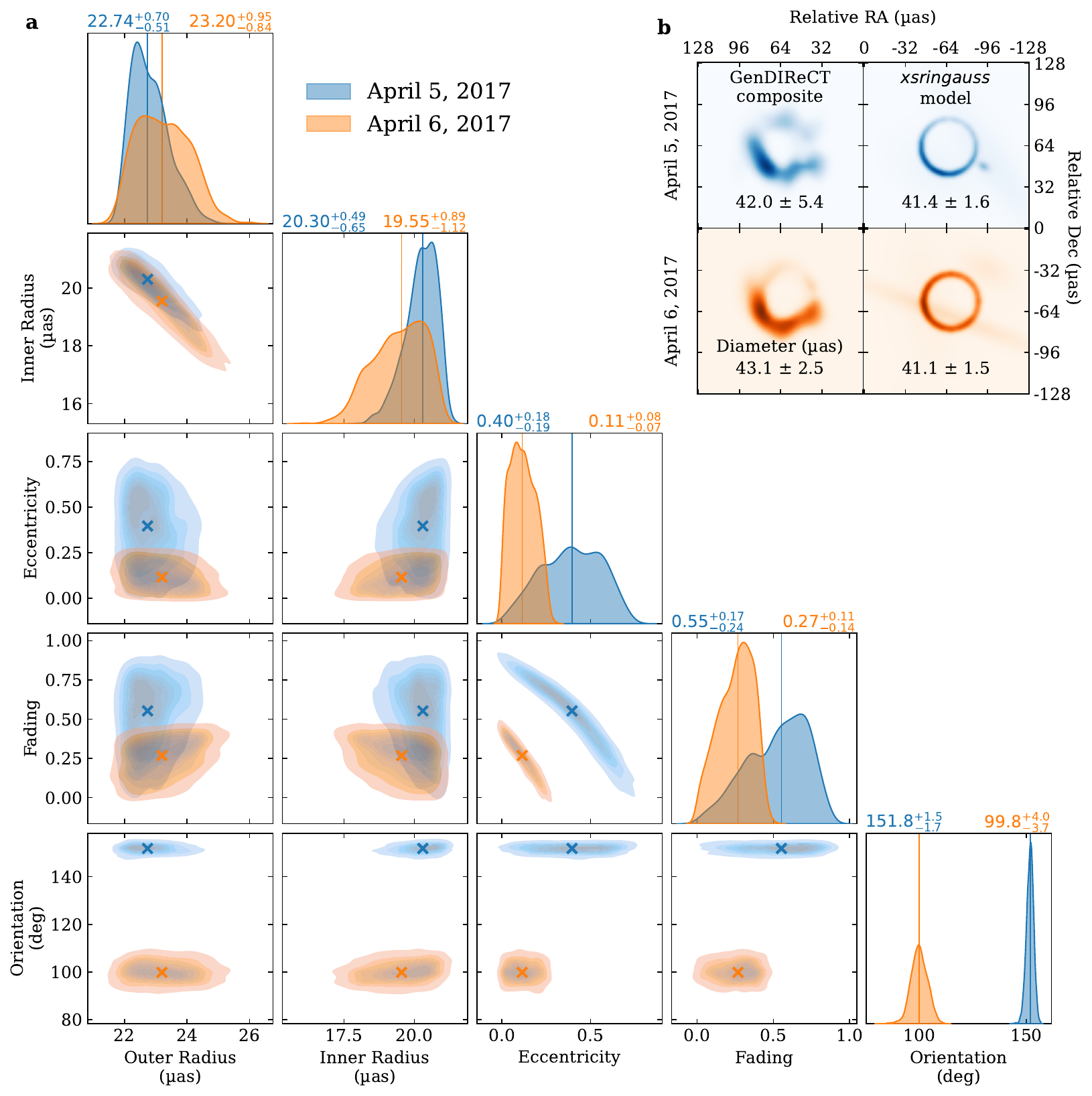}}
\end{center}
\caption{Similar to Figure \ref{fig:corner-main}, but for the April 5, 2017 and April 6, 2017 datasets.}
\label{fig:corner56}
\end{figure}

\begin{figure}[htb]
\begin{center}
\resizebox{16cm}{!}{\includegraphics[trim={0 0 0 0},clip]{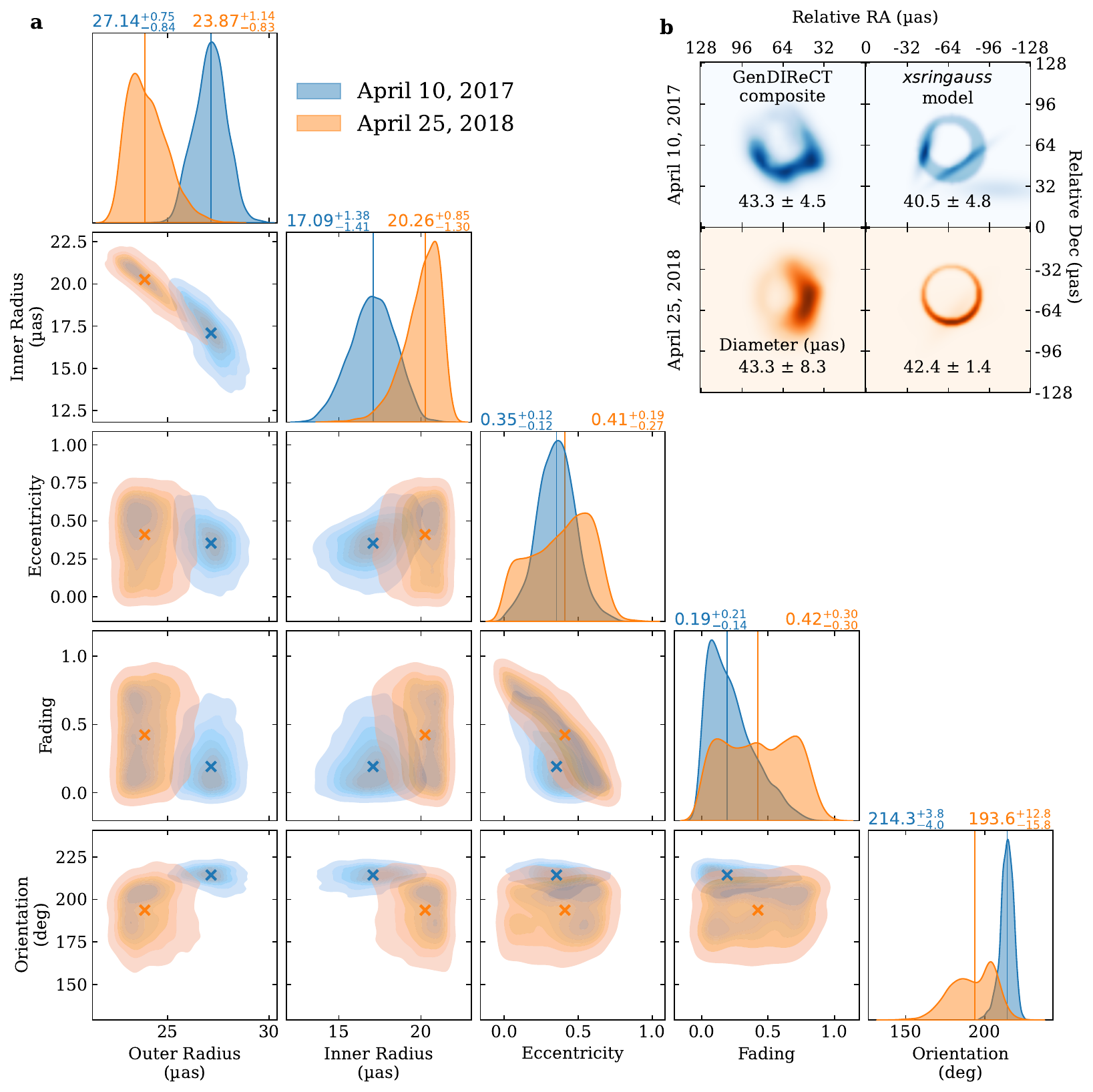}}
\end{center}
\caption{Similar to Figure \ref{fig:corner-main}, but for the April 10, 2017 and April 25, 2018 datasets.}
\label{fig:corner1015}
\end{figure}

\begin{figure}[htb]
\begin{center}
\resizebox{16cm}{!}{\includegraphics[trim={0 0 0 0},clip]{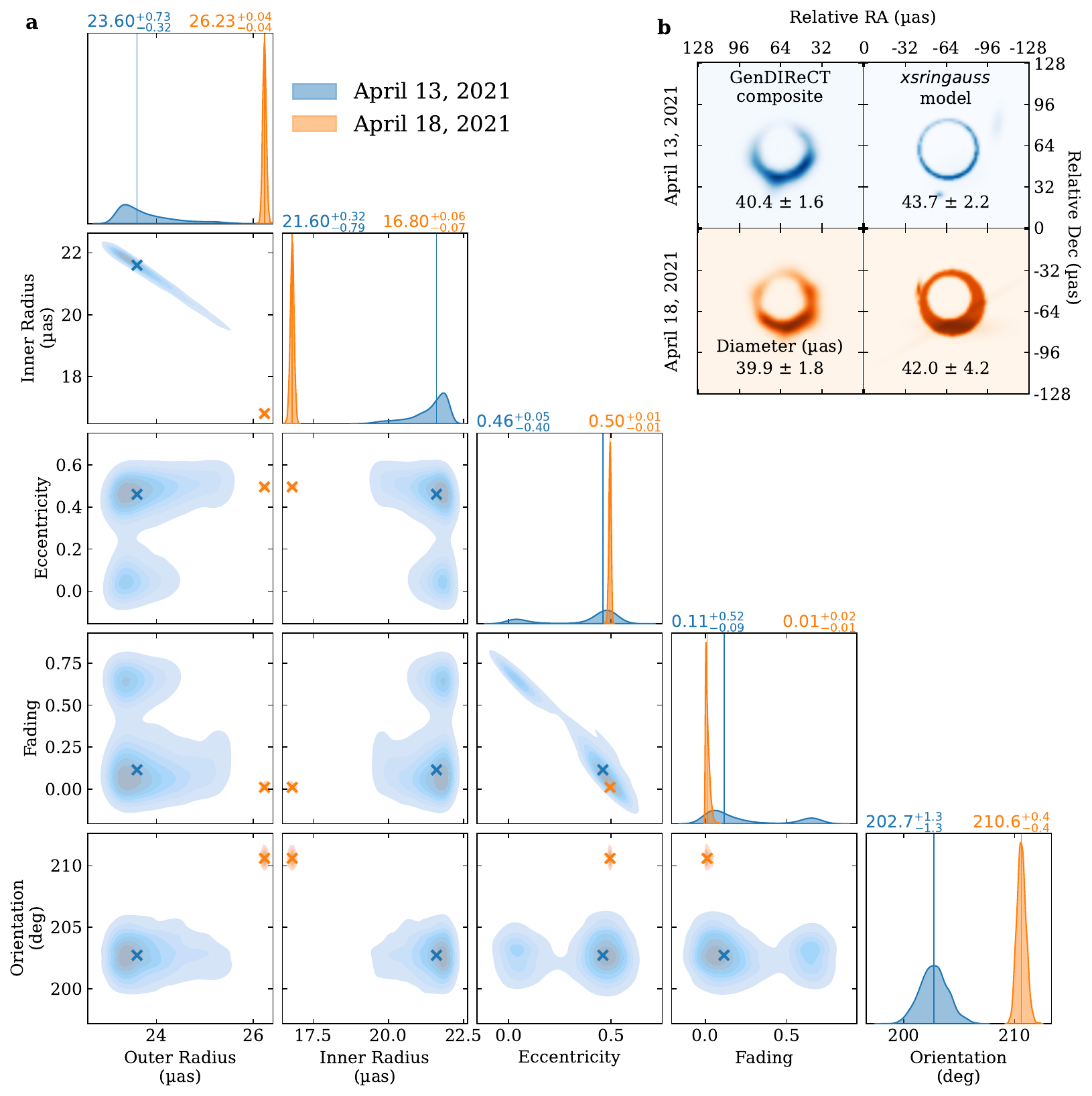}}
\end{center}
\caption{Similar to Figure \ref{fig:corner-main}, but for the April 13, 2021 and April 18, 2021 datasets.}
\label{fig:corner1318}
\end{figure}

Each image component can be further independently augmented by translation, rotation, and Gaussian blur. In general, image translations are represented with an additional phase term in the Fourier domain, $e^{i2\pi(l_0 u + m_0 v)}$, where $(l_0,m_0)$ denotes the translation in $(l,m)$ coordinates. Image rotation is modeled by a rotation of the $uv$-coordinates. Then, by the convolution theorem, blurring in the image domain with a circular Gaussian kernel is a point-wise multiplication in the Fourier domain by $e^{-i2\pi^2\sigma_b^2\rho^2}$, where $\sigma_b$ is the width of the blurring kernel. The final image model and its corresponding analytic Fourier transform is described by 28 parameters. 
Throughout this work, the \xsrg\ model is fit to observed data with \dynesty, a Bayesian dynamic nested sampling algorithm, adopting uniform priors for all parameters. The final fitted image model is formed from the median value of the one-dimensional marginalised posteriors for each parameter. The \xsrg\ model parameter covariance plot for the highest-quality datasets of 2017 and 2018 is presented in Figures \ref{fig:corner-main} of the main text, and we present covariance plots for all other datasets in Extended Data Figures \ref{fig:corner56}, \ref{fig:corner1015}, and \ref{fig:corner1318}. We then apply the image-domain feature extraction analysis on samples from the posterior to measure all characteristic ring parameters (Extended Data Table \ref{tab:GCFit-ring-features}).

\subsection*{Inferring the black hole mass}
Inferring the black hole's gravitational radius from the measured ring diameter requires independently calibrating each imaging method with synthetic data from simulated general-relativistic magnetohydrodynamic models of lensed black hole accretion emission\cite{EHT_2019_ShadowMass}. One common simplification is that the observed size of the reconstructed ring should follow a simple scaling relationship with the gravitational radius, $d_{\rm ring} = \alpha\theta_g$. By calibrating to a large ensemble of simulations with different black hole spins, the EHTC determined the calibration factor, $\alpha$, from the image-domain feature extraction method to be between $10.67-11.01$, where the listed range is due to variations from different image reconstruction algorithms. For geometric modelling methods, the $\alpha$ calibration factor ranges between $11.01-12.06$. The median of all $\alpha$ calibrations is $\alpha\simeq 11.0$, which is the calibration factor adopted by the EHTC for methods that have not been independently calibrated\cite{EHT_2024_M87_PerShad}. Therefore, we also provisionally apply $\alpha\sim11.0$ as our calibration factor in this study.

The associated theoretical uncertainty in the calibrated scaling factor, which arises from recognising the dataset as a single sample from a dynamic system with unknown physical properties, is $\sigma_\alpha = 0.4$\cite{EHT_2019_ShadowMass}. In this work, we also adopt the mean \mes\ distance measure from the EHTC, which is $16.8^{+0.8}_{-0.7}\,{\rm Mpc}$\cite{EHT_2019_Imaging} based on studies of high-resolution resolved stellar distribution fluctuations\cite{Blakeslee_2009, Cantiello_2018} and the tip of the red giant branch standard candle method\cite{Bird_2010}.

The prior expectation for the mass of the central black hole in M87 is primarily sourced from high-quality gas\cite{Walsh_2013} and stellar\cite{Gebhardt_2011} dynamical estimates, which measured $M_{\rm gas} = 3.5^{+0.9}_{-0.7} \times10^9\, M_{\odot}$ and $M_{\rm star} = (6.6 \pm0.4)\times10^9\,M_{\odot}$, respectively. The corresponding gravitational radii are $\theta_{\rm gas} = 2.1^{+0.5}_{-0.4}\,\mu{\rm as}$ and $\theta_{\rm star} = 3.9\pm0.3\,\mu{\rm as}$ if we adopt a distance measure of $16.8^{+0.8}_{-0.7}\,{\rm Mpc}$. Using the median $\alpha=11.0\pm0.4$ calibration factor and theoretical uncertainty, the prior ring diameters are $d_{\rm gas} = 22.6^{+5.9}_{-4.7}\,\mu{\rm as}$ and $d_{\rm star} = 42.7^{+3.5}_{-3.6} \,\mu{\rm as}$. We therefore estimate a black hole mass of $M_{\rm BH} = 6.5^{+0.5}_{-0.4} (11.0/\alpha) \times 10^9\, M_{\odot}$ with \gendirect's independently measured mean ring diameter. The full range of $\alpha=10.67-12.06$ calibration factors implies an estimated black hole mass in the range $M_{\rm BH} = (6.0-6.7) \times 10^9\, M_{\odot}$. Based on this, we conservatively estimate our black hole mass as $M_{\rm BH} = (6.5 \pm 0.5) \times 10^9\, M_{\odot}$.

\subsection*{Synthetic \mes\ images}
In order to compare the reconstructed composites to synthetic \mes\ images, we retrieve publicly available general relativistic magnetohydrodynamic (GRMHD) data\footnote{\href{https://thz.astro.illinois.edu/resources.html}{https://thz.astro.illinois.edu/resources.html}} from the University of Illinois survey\cite{Dhruv_2025_illinoisGRMHD}, which includes fluid data for 1000 timesteps (in $R_g/c$) in the quasi-stationary equilibrium state and five black hole spin prescriptions ($a_{*} = -0.94, -0.5, 0, +0.5, +0.94$), evolving under the magnetically arrested disk (MAD) or standard and normal evolution (SANE) regime. In this work, we consider only the MAD fluid models, which is the more likely evolutionary scenario for \mes\ based on jet power constraints\cite{EHT_2019_Vphysical}.

Synthetic 230 GHz images of \mes\ are produced using the general relativistic radiative transfer \ipole\ algorithm\cite{Moscibrodzka_2018_ipole} for all timesteps, spins, and inclination angles between $5^{\circ}$ and $90^{\circ}$ with a step size of $1^{\circ}$. Note that we adopt a coupled electron temperature model characterised by $R_{\rm high} = 40$, which determines the relationship between the ion and electron temperature. Though $R_{\rm high}$ is a free parameter, the time-averaged MAD images are nearly independent of $R_{\rm high}$ and are sensitive primarily on black hole spin for a fixed inclination\cite{EHT_2019_Vphysical}. In this work, we compare all beam-convolved synthetic images to the reconstructed \mes\ composites in the image-domain using the normalised cross-correlation metric after correcting for the observer position angle and making linear adjustments to spatial scale, where both the rotation and scale adjustments of the synthetic images are chosen independently for each parameter combination of spin, inclination, and timestep, to maximise the resulting normalised cross-correlation comparison with the composite reconstructions. Simulations in ideal GRMHD are scale-agnostic, enabling such adjustments to be made without affecting the underlying physics\cite{Gammie_2003_HARM, Dhruv_2025_illinoisGRMHD}. We also repeat our analysis for $R_{\rm high} = 80$ , presenting the figure in the \gendirect\ repository, and find a mild $\sim2^{\circ}$ shift toward lower inclinations.

\subsection*{Data availability} 
The science-ready observational data underlying this paper are publicly available from the Event Horizon Telescope Collaboration website\footnote{\href{https://eventhorizontelescope.org/}{https://eventhorizontelescope.org/}}, namely EHT Data Product Codes \href{https://dc.cyverse.org/dataset/first_m87_eht_results_calibrated_data}{2019-D01-01}, \href{https://doi.org/10.25739/epm5-r371}{2024-D01-01}, and \href{https://datacommons.cyverse.org/browse/iplant/home/ckc/EHTC_M87-2018+2021_Jul2026}{2026-D01-01}. Post-processed data products or image reconstructions can be located in the \gendirect\ code repository.

\subsection*{Code availability} 
Code for \gendirect\cite{GenDIReCT_Zenodo}, geometric modelling with image-domain feature extraction\cite{BASIS_Zenodo}, and the closure invariants formalism\cite{CI_Package} are publicly available.

\section*{Acknowledgements}
Software packages used in this study include \textsc{Numpy}\cite{Numpy_2011}, \textsc{Scipy}\cite{Scipy_2020}, \textsc{PyTorch}\cite{PyTorch}, \ehtim{\cite{Chael_2018_ehtim}}, \gendirect{\cite{GenDIReCT_Zenodo}}, \textsc{BASIS}\cite{BASIS_Zenodo}, \textsc{ClosureInvariants}\cite{CI_Package}, \dynesty{\cite{dynesty}}, \bilby{\cite{Ashton_2019_bilby}}, \ipole{\cite{Moscibrodzka_2018_ipole}}, and \textsc{Matplotlib}\cite{Matplotlib_2007}. We acknowledge use of high performance computing resources from the Commonwealth Scientific and Industrial Research Organisation. Inputs from Rajaram Nityananda and Ron Ekers are gratefully acknowledged.

\section*{Author contributions statement}
S.~Lai developed \gendirect, \textsc{BASIS}, executed the analysis from Event Horizon Telescope data to reconstructed image, and drafted most of the manuscript. N.~Thyagarajan produced the closure invariants formalism and associated software package \textsc{ClosureInvariants} used by \gendirect\ for imaging, and made contributions to the analysis and the manuscript. O. I.~Wong and F.~Diakogiannis provided support on machine learning techniques which significantly contributed to the development of \gendirect\ and provided constructive feedback on the manuscript. 

\section*{Funding}
The authors declare that no relevant external funding was received for this work. 

\section*{Competing interests} 
The authors declare no competing interests.

\newpage

\begin{table}[h]
\begin{center}
\caption {\label{tab:GCFit-ring-features} Similar to Table \ref{tab:morph-features}, but with ring properties measured from geometric modelling of the closure invariants datasets from each independent epoch. For all parameters, we present the mean and standard deviation derived from sampling the posterior and applying image-domain feature extraction.}
\resizebox{\textwidth}{!}{
\begin{tabular}{ll|ccccc}
\hline \hline
Date &  & Ring diameter & Ring width & Orientation & Azimuthal asymmetry & Central brightness fraction \\
& & $2\bar r_{\rm pk}$ [$\mu{\rm{as}}$] & $w_r$ [$\mu${\rm{as}}] & $\eta$ [$^{\circ}$] & $A$ & $f_c$ \\ \hline
2017 & April 5 & $41.5 \pm 0.1$ & $5.0 \pm 0.4$ & $165.7 \pm 1.3$ & $0.262 \pm 0.003$ & $0.19 \pm 0.12$ 
\\
& April 6 & $40.9 \pm 0.1$ & $5.2 \pm 0.2$ & $87.2 \pm 2.0$ & $0.135 \pm 0.003$ & $0.12 \pm 0.08$ 
\\
& April 10 & $40.5 \pm 0.2$ & $8.5 \pm 0.1$ & $166.1 \pm 1.0$ & $0.160 \pm 0.002$ & $0.13 \pm 0.07$ 
\\ 
& April 11 & $43.0 \pm 0.1$ & $6.1 \pm 0.1$ & $167.7 \pm 0.7$ & $0.284 \pm 0.002$ & $0.18 \pm 0.07$ 
\\ \hline
2018 & April 21 & $43.4 \pm 0.1$ & $5.8 \pm 0.1$ & $204.3 \pm 0.5$ & $0.257 \pm 0.003$ & $0.12 \pm 0.07$ 
\\
& April 25 & $42.2 \pm 0.1$ & $5.5 \pm 0.3$ & $190.5 \pm 0.9$ & $0.282 \pm 0.003$ & $0.19 \pm 0.13$ 
\\ \hline
2021 & April 13 & $43.9 \pm 0.2$ & $5.0 \pm 0.6$ & $184.8 \pm 1.9$ & $0.183 \pm 0.003$ & $0.28 \pm 0.08$
 \\
& April 18 & $42.7 \pm 0.2$ & $8.1 \pm 0.2$ & $195.6 \pm 0.6$ & $0.179 \pm 0.005$ & $0.17 \pm 0.01$
\\ \hline
Mean & & $42.3 \pm 1.1$ & -- & -- & --& --
\\
\hline \hline
\end{tabular}
}
\end{center}
\end{table}


\end{document}